\documentclass[preprintnumbers,showpacs,showkeys,preprint,secnumarabic,amssymb,amsmath,amsfonts,aps, pre]{revtex4-1}

\usepackage{calrsfs}
\usepackage{float}
\usepackage{amsmath}
\usepackage{amsfonts}
\usepackage{amssymb}

\usepackage{textcomp}
\everymath{\displaystyle}
\DeclareMathAlphabet{\pazocal}{OMS}{zplm}{m}{n}

\usepackage{makeidx}
\usepackage{gensymb}
\usepackage{ragged2e}
\usepackage[pdftex]{graphicx, color}
\usepackage[dvipsnames]{xcolor}
\usepackage{float}
\usepackage{here}
\usepackage{color}
\usepackage{hyperref}
\newcommand{\affA}{Institute of Neuroscience and Medicine INM-9, and Institute for Advanced Simulation IAS-5, Forschungszentrum J\"ulich, 52428 J\"ulich, Germany}
\newcommand{\affB}{ Department of Physics, Faculty of Mathematics, Computer Science and Natural Sciences, Aachen University, 52062 Aachen, Germany}
\newcommand{\affC}{Physics and Materials Science Research Unit, University of Luxembourg, L-1511
Luxembourg, Luxembourg}
\newcommand{\affD}{Dipartimento di Fisica Universit\`a di Firenze, and
I.N.F.N., Sezione di Firenze, via G. Sansone 1, I-50019 Sesto Fiorentino, Italy}
\newcommand{\affE}{Aix-Marseille University, Marseille, France}
\newcommand{\affF}{CNRS Centre de Physique Th\'eorique UMR7332,
13288 Marseille, France}

\begin{document}
\title{Hamiltonian chaos and differential geometry of configuration space-time} 

\author{Loris Di Cairano}
\email{l.di.cairano@fz-juelich.de}
\affiliation{\affA}\affiliation{\affB}

\author{Matteo Gori}
\email{gori6matteo@gmail.com}
\affiliation{\affC}

\author{Giulio Pettini}
\email{pettini@fi.unifi.it}
\affiliation{\affD}

\author{Marco Pettini}
\email{pettini@cpt.univ-mrs.fr}
\affiliation{\affE}\affiliation{\affF}

\date{\today}
\begin{abstract}
This paper tackles Hamiltonian chaos by means of elementary tools of Riemannian geometry. More precisely, a Hamiltonian flow is identified with a geodesic flow on configuration space-time endowed with a suitable metric due to Eisenhart. Until now, this framework has never been given attention to describe chaotic dynamics. A gap that is filled in the present work. In a Riemannian-geometric context, the stability/instability of the dynamics depends on the curvature properties of the ambient manifold and is investigated by means of the Jacobi--Levi-Civita (JLC) equation for geodesic spread. It is confirmed that the dominant mechanism at the ground of chaotic dynamics is parametric instability due to curvature variations along the geodesics. A comparison is reported of the outcomes of the JLC equation written also for the Jacobi metric on configuration  space and for another metric due to Eisenhart on an extended configuration space-time. This has been applied to the H\'enon-Heiles model, a two-degrees of freedom system. Then the study has been extended to the 1D classical Heisenberg XY model at a large number of degrees of freedom. Both the advantages and drawbacks of this geometrization of Hamiltonian dynamics are discussed. Finally, a quick hint is put forward concerning the possible extension of the differential-geometric investigation of chaos in generic dynamical systems, including dissipative ones, by resorting to Finsler manifolds.  

\end{abstract}
\maketitle

\section{Introduction}
As is well known, a generic property of nonlinear dynamical systems, described by a system of differential equations, is the presence of \textit{deterministic chaos}. This means that despite the deterministic nature of a dynamical system of this kind, that is, despite the Cauchy's theorem of existence and unicity of the solutions of a system of differential equations, the property of \textit{predictability} of the dynamics for arbitrary times is lost in the absence of \textit{stability} of the dynamics \cite{chaos,wiggins,chaos1}. Such a dramatic consequence of the breaking of integrability of a three body problem was already pointed out by Poincar\'e while describing the complexity of the homoclinic tangles in the proximity of hyperbolic points in phase space \cite{poincare}. It was at the beginning of the 60's of the last century that for the first time the consequences of homoclinic tangles in phase space of a nonlinear Hamiltonian system became visually evident. This was thanks to the numerical integration of the equations of motion of the celebrated H\'enon-Heiles model \cite{henon}. The numerically worked out surfaces of section in phase space displayed what Poincar\'e declared to be unable even to dare to attempt drawing \cite{poincare}.
For many decades now, a huge amount of work has been done, both numerical and mathematical, on deterministic chaos. However, especially for many degrees of freedom systems, a theoretical explanation of the origin of chaos has been lacking. Homoclinic intersections certainly provide an elegant explanation of the origin of chaos in both dissipative and Hamiltonian systems, but apply to 1.5 or two degrees of freedom systems. Beautiful theorems on Axiom A systems \cite{chaos} and Anosov flows \cite{anosov} cannot account for the emergence of chaos in dynamical systems of physical relevance.
An independent attempt to explain the origin of chaos in Hamiltonian systems was put forward by N.S.Krylov who resorted to the possibility of identifying a Hamiltonian flow with a geodesic flow in configuration space to try to explain the origin of the dynamical
instability (which we nowadays call deterministic chaos) that could explain the spontaneous tendency to thermalization of many body systems. Krylov's pioneering approach focused on the search for negative curvatures in configuration space equipped with a suitable metric \cite{krylov}. Krylov's work inspired abstract ergodic theory but did not go too far to explain the origin of chaos in Hamiltonian dynamical systems. For instance, in the case of the already mentioned H\'enon-Heiles model, it turns out that no region of negative curvature can be found in configuration space, therefore Krylov's intuition has been discarded for a long time. However, more recently, on the basis of numerical "experiments" it has been shown that chaos in Hamiltonian flows of physical relevance stems from another mechanism, parametric instability, which will be discussed throughout this paper.
The Riemannian-geometric approach to explaining the origin of chaos in Hamiltonian flows is based on two fundamental elements \cite{marco}: \textit{i)} the identification of a Hamiltonian flow with a geodesic flow of a Riemannian manifold equipped with a suitable metric, so that the geodesic equations
\begin{equation}
\frac{d^2 q^i}{ds^2}+ \Gamma^i_{jk}\frac{dq^j}{ds}\frac{dq^k}{ds}=0~.
\label{geod-mot}
\end{equation}
coincide with Newton's equations
\begin{equation}
\frac{d^2 q^i}{dt^2} = - \frac{\partial V(q)}{\partial q^i}~.
\label{newton}
\end{equation}
a Hamiltonian flow - of which the kinetic energy is a quadratic form in the velocities, that is, $H = \frac{1}{2} a_{ik} p^i p^k + V(q_1,\ldots,q_N)$ - is equivalent to the solutions of Newton’s equations of motion stemming from a Lagrangian function $L = \frac{1}{2} a_{ik} \dot q^i\dot q^k- V(q_1,\ldots,q_N)$;

\noindent \textit{ii)} the description of the stability/instability of the dynamics by means of the Jacobi--Levi-Civita (JLC) equation for the geodesic spread measured by the geodesic deviation vector field $J$ (which locally measures the
distance between nearby geodesics), which in a parallel-transported
frame reads
\begin{equation}
\frac{d^2 J^k}{ds^2} +
R^k_{~ijr} \frac{dq^i}{ds}{J^j}\frac{dq^r}{ds} = 0~.
\label{geod-dev}
\end{equation}
where $R^k_{~ijr}$ are the components of the Riemann-Christoffel curvature tensor. 

The most natural geometrization of Hamiltonian dynamics in a Riemannian framework \footnote{The natural and elegant geometric setting of Hamiltonian dynamics is provided by symplectic geometry. This geometrical framework is very powerful to study, for example, symmetries.
However, symplectic manifolds are not endowed with a metric, and without a metric we do
not know how to measure the distance between two nearby phase space trajectories and thus
to study their stability/instability through the time evolution of such a distance.} is a consequence of Maupertuis least action principle for isoenergetic paths
\begin{equation}
\delta \ \int_{q(t_0)}^{q(t_1)}\ d t\ W(q, \dot q) = 0\ ,
\label{Maper}
\end{equation}
where $W(q, \dot q)=\{[E- V(q)]a_{ik} {\dot q}^i{\dot q}^i\}^{1/2}$,
which is equivalent to the variational definition of a {\it geodesic}
line  on a Riemannian manifold, a line of stationary or minimum length
joining the points $A$ and $B$:
\begin{equation}
\delta \ \int_{A}^{B}\ d s = 0\ .
\label{geode}
\end{equation}
If the subset of configuration space $M_E=\{(q_1,\ldots,q_N)\in{\mathbb R}^N \vert V(q_1,\ldots,q_N)<E\}$ is given the non-Euclidean metric of components 
\begin{equation}
g_{ij} =2 [E- V(q)]a_{ik}\ ,
\label{metri}
\end{equation}
whence the infinitesimal arc element $ds^2= 4[E- V(q)]^2 dq_i\ dq^i$, then
Newton's equations \eqref{newton} are retrieved from the geodesic equations \eqref{geod-mot}.

The JLC equation for the geodesic spread can be rewritten as \cite{book}
\begin{equation}
\frac{d^2J^k}{ds^2}+2\Gamma^k_{ij}\frac{dq^i}{ds} \frac{d J^j}{ds} +
\left( \frac{\partial\Gamma^k_{ri}}{\partial q^j}\right)\,\frac{dq^r}{ds}
\frac{dq^i}{ds}\,{J^j} = 0\ ,
\label{jlce}
\end{equation}
which has general validity {\it independently} of the metric of the ambient manifold.

Importantly, there are other Riemannian manifolds, endowed with different metric tensors, to geometrize Hamiltonian dynamics \cite{book}.
Two of these alternatives are concisely described in the following. One brings about the standard tangent dynamics equation as geodesic spread (JLC) equation, whereas the second one has never been investigated hitherto to describe chaos in Hamiltonian flows. This gap is filled in the present work. The choice among these manifolds is driven by practical computational reasons as will be discussed in what follows.

\section{Eisenhart  Geometrization of Hamiltonian dynamics}
\label{section1}
It is worth summarizing some basic facts of a geometrization of Hamiltonian dynamics which makes a direct and unexpected link between the standard tangent dynamics equations, used to numerically compute Lyapunov exponents, and the JLC equation for the geodesic spread \cite{book}.
\subsection{Eisenhart Metric on Enlarged Configuration Space-Time \texorpdfstring{$M\times\mathbb{R}^2$}{MxR2}}
\label{Eisenhart_metric2}
L.P.Eisenhart proposed a geometric formulation of Newtonian dynamics that
makes use, as ambient space, of an enlarged configuration space-time
$M\times \mathbb{R}^2$ of local coordinates
$(q^0,q^1,\ldots,q^i,\ldots,q^N,q^{N+1})$. This space can be
endowed
with a nondegenerate pseudo-Riemannian metric \cite{Eisenhart} whose
arc length is
\begin{equation}
ds^2 = \left( g_e \right)_{\mu\nu}\, dq^{\mu}dq^{\nu} =
a_{ij} \, dq^i dq^j -2V(q)(dq^0)^2
+ 2\, dq^0 dq^{N+1} ~,
\label{g_E}
\end{equation}
where $\mu$ and $\nu$ run from $0$ to $N+1$ and  $i$ and $j$ run from 1 to $N$.
The relation between the geodesics of
this manifold and the natural motions of the dynamical system
is contained in the following theorem \cite{lichnerowicz}:

\textbf{Theorem.}
\textit{The natural motions of a Hamiltonian dynamical system
are obtained as the canonical projection
of the geodesics
of $(M\times \mathbb{R}^2,g_e)$ on the configuration
space-time,
$\pi : M\times \mathbb{R}^2 \mapsto M\times \mathbb{R}$.
Among the totality of geodesics, only those whose
arc lengths are
positive definite and are given by
\begin{equation}
ds^2 = c_1^2 dt^2
\label{ds_Eisenhart}
\end{equation}
correspond to natural motions; the
condition (\ref{ds_Eisenhart}) can be equivalently cast in the following
integral form
as a condition on the extra coordinate $q^{N+1}$:
\begin{equation}
q^{N+1} = \frac{c_1^2}{2} t + c^2_2 - \int_0^t { L}\,
d\tau~,
\label{qN+1}
\end{equation}
where $c_1$ and $c_2$ are given real constants.
Conversely, given a point $P \in M \times \mathbb{R}$ belonging
to a trajectory
of the system, and given two constants $c_1$ and $c_2$, the
point
$P' = \pi^{-1} (P) \in M \times \mathbb{R}^2$, with
$q^{N+1}$ given by
(\ref{qN+1}), describes a geodesic curve in $(M\times \mathbb{R}^2,g_e)$
such that $ds^2 = c_1^2 dt^2$.   }

\noindent For the full proof, see \cite{lichnerowicz}.
Since the constant $c_1$ is arbitrary, we will always set
$c_1^2 = 1$ in order that $ds^2 = dt^2$ on the physical geodesics.

From (\ref{g_E}) it follows that
the explicit table of the components of the Eisenhart metric is given by
\begin{equation}
g_e = \left(
\begin{array}{ccccc}
-2V(q)& 0       & \cdots        & 0     & 1     \\
0       & a_{11}& \cdots        & a_{1N}& 0     \\
\vdots  & \vdots& \ddots        & \vdots& \vdots\\
0       & a_{N1}& \cdots        & a_{NN}& 0     \\
1       & 0     & \cdots        & 0     & 0     \\
\end{array} \right)\ , \label{gE}
\end{equation}
where $a_{ij}$ is the kinetic energy metric. The Christoffel coefficients
\begin{equation}
\Gamma^i_{jk}=\frac{1}{2}g^{im}\left( \frac{\partial g_{mk}}{\partial q^j}
+\frac{\partial g_{mj}}{\partial q^k}-\frac{\partial g_{jk}}{\partial q^m}
\right) 
\label{christoff}
\end{equation}
for $g_e$ and with $a_{ij} = \delta_{ij}$ are found to be non-vanishing only in the following cases
\begin{equation}
\Gamma^i_{00} = - \Gamma^{N+1}_{0i} = \partial_i V~,
\label{Gamma_E}
\end{equation}
where $\partial_i =\partial /\partial q^i$ so that the geodesic equations read
\begin{eqnarray}
\frac{d^2q^0}{ds^2}  &=&  0 ~, \label{eqgeo0}\\
\frac{d^2q^i}{ds^2} +\Gamma^i_{00}
\frac{dq^0}{ds}\frac{dq^0}{ds}  &=&  0\ ,
\label{eqgeoi}\\
\frac{d^2q^{N+1}}{ds^2} +\Gamma^{N+1}_{0i} \frac{dq^0}{ds}
\frac{dq^i}{ds}  &=&  0 \ ; \label{eqgeoN+1}
\end{eqnarray}
using $ds = dt$ one obtains
\begin{eqnarray}
\frac{d^2q^0}{dt^2}  &=&  0\ , \label{eqgeo0t}\\
\frac{d^2q^i}{dt^2}  &=&  - \frac{\partial V}{\partial q_i}
~,\label{eqgeoit}\\
\frac{d^2q^{N+1}}{dt^2} & =&   - \frac{d{ L}}{dt} ~.
\label{eqgeoN+1t}
\end{eqnarray}
Equation (\ref{eqgeo0t}) states only that $q^0=t$.
The $N$ equations (\ref{eqgeoit})
are Newton's equations, and (\ref{eqgeoN+1t}) is
the differential version of (\ref{qN+1}).

The fact that in the framework of the Eisenhart metric the dynamics can be
geometrized with an affine parametrization of the arc length, i.e.,
$ds = dt$, will be extremely useful in the following, together
with the remarkably simple curvature properties of the Eisenhart metric. 
\subsubsection{Curvature of \texorpdfstring{$(M\times\mathbb{R}^2,g_e)$}{MxR2,ge}}
The curvature properties of the Eisenhart metric $g_e$ are much
simpler than those
of the Jacobi metric, and this is obviously a great advantage from a
computational point of view. The components of the Riemann--Christoffel curvature tensor are
\begin{equation}
R^k_{~ijr}  =
\left(
\Gamma^t_{ri}\Gamma^k_{jt}-\Gamma^t_{ji}\Gamma^k_{rt} +\partial_j\Gamma^k_{ri}
-\partial_r\Gamma^k_{ji}\right)\ .
\label{Curv-tens}
\end{equation}
Hence, and after Eq.\eqref{Gamma_E}, the only non-vanishing
components of the  curvature tensor are 
\begin{equation}
R_{0i0j} = \partial_i\partial_j V
\end{equation}
hence the Ricci tensor has only one nonzero 
component  
\begin{equation}
R_{00} = \triangle V 
\label{ricci_eisenhart}
\end{equation}
so that the Ricci 
curvature is 
\begin{equation}
K_R(q,\dot q)=R_{00}\dot q^0\dot q^0\equiv \triangle V\ ,
\label{Ricci-ge}
\end{equation}
and the scalar curvature is identically vanishing ${\mathcal R}(q) = 0~.$

\subsubsection{Geodesic Spread Equation for the Eisenhart Metric \texorpdfstring{$g_e$}{ge}}
The Jacobi equation (\ref{geod-dev}) for $(M\times\mathbb{R}^2,g_e)$ takes the form 
\begin{eqnarray}
\frac{\nabla^2 J^0}{ds^2} +R^0_{i0j}\frac{dq^i}{ds}J^0
\frac{dq^j}{ds}+R^0_{0ij}\frac{dq^0}{ds}J^i\frac{dq^j}
{ds}&=& 0 \ ,~~~\label{JLC_gE_1}\\
\frac{\nabla^2 J^i}{ds^2} +R^i_{0j0}\left( \frac{dq^0}{ds}\right)
^2J^j+R^i_{00j}\frac{dq^0}{ds}J^0\frac{dq^j}{ds}+
R^i_{j00}\frac{dq^j}{ds}J^0\frac{dq^0}{ds}&=&0\ ,~~~
\label{JLC_gE_2}\\
\frac{\nabla^2 J^{N+1}}{ds^2}
+R^{N+1}_{i0j}\frac{dq^i}{ds}J^0
\frac{dq^j}{ds}+R^{N+1}_{ij0}\frac{dq^i}{ds}J^j\frac{dq^0}
{ds}&=&0\ ,~~~ \label{JLC_gE_3}
\label{eq_JLC_gE}
\end{eqnarray}
and since $\Gamma^0_{ij}=0$ and $\Gamma^i_{0k} = 0$ it is  $\nabla J^0/ds = dJ^0/ds$, $R^0_{~ijk}=0$, and ${\nabla J^i}/{ds} = {d J^i}/{ds}$, the only accelerating components of the vector field $J$ are found to obey the 
equations
\begin{equation}
\frac{d^2 J^i}{ds^2} + \frac{\partial^2 V}{\partial q_i
\partial q^k}
\left(\frac{dq^0}{ds}\right)^2 J^k = 0\ .
\label{eq_jacobi_E}
\end{equation}
and using $dq^0/ds = 1$ one is left with
\begin{equation}
\frac{d^2J^i}{dt^2}+\frac{\partial^2V}{\partial q_i\partial q^k}\ J^k = 0\ ,
\label{dyntanT}
\end{equation}
the usual tangent dynamics equations. This fact is a crucial point in the development of a geometric theory of
Hamiltonian chaos because there is no new definition of chaos in the geometric context. In fact, the numerical Lyapunov
exponents computed by means of Eqs.\eqref{dyntanT} already belong to geometric treatment of chaotic geodesic flows.

\subsection{Eisenhart Metric on Configuration Space-Time \texorpdfstring{$M\times\mathbb{R}$}{MxR}}
\label{Eisenhart_metric}
Another interesting choice of the ambient space and Riemannian metric to reformulate
Newtonian dynamics in a geometric language was also proposed by Eisenhart \cite{Eisenhart}.
If and how the description of Hamiltonian chaos in this framework is coherent with the results obtained by standard treatment 
based on the tangent-dynamics/JLC equations discussed in the preceding section has never been investigated before. 

This geometric formulation makes use of an enlarged configuration space $M\times \mathbb{R}$, with local
coordinates $(q^0,q^1,\ldots,q^N)$, where a proper Riemannian metric $G_e$ is
defined to give
\begin{equation}
ds^2 = \left(G_e \right)_{\mu\nu}\, dq^{\mu}dq^{\nu} =
a_{ij} \, dq^i dq^j +  A(q)\, (dq^0)^2 ~,
\label{g1_E}
\end{equation}
where $\mu$ and $\nu$ run from $0$ to $N$ and  $i$ and $j$ run from 1 to $N$,
and the function $A(q)$ does not explicitly depend on time. With the choice
$1/[2A(q)]= V(q)+\eta$ and under the condition
\begin{equation}
q^{0} =  2 \int_0^t V(q)\, d\tau + 2\eta t\ ,
\label{q(N+1)}
\end{equation}
for the extra variable it
can easily be seen that the geodesics of the manifold $(M\times\mathbb{R}, G_e)$
are the natural motions of standard autonomous Hamiltonian systems.
Since $\frac{1}{2}a_{ij}\dot q^i\dot q^j + V(q) = E$,
where $E$ is the energy constant along a geodesic, we can see that the following relation
exists between $q^0$ and the action:
\begin{equation}
q^{0} = - 2 \int_0^t T\, d\tau + 2(E + \eta) t\ .
\label{q0action}
\end{equation}

\noindent Explicitly, the metric $G_e$ reads as
\begin{equation}
G_e = \left(
\begin{array}{cccc}
[2V(q)+2\eta]^{-1}& 0       & \cdots        & 0   \\
0       & a_{11}& \cdots        & a_{1N}   \\
\vdots  & \vdots& \ddots        & \vdots\\
0       & a_{N1}& \cdots        & a_{NN}     \\
\end{array} \right)\ , \label{g1E}
\end{equation}
and together with the condition (\ref{q0action}), this gives an affine
parametrization of the arc length with the physical time, i.e.,
$ds^2=2 ( E + \eta) dt^2$, along the geodesics that coincide with natural
motions. The constant $\eta$ can be set equal to an arbitrary
value greater than the largest value of $\vert E\vert$ so that the metric
$G_e$ is nonsingular.
This metric is a priori very interesting because it seems to have
some better property than the Jacobi metric and than the previous metric $g_e$. 
In fact, at variance with the Jacobi metric $g_J$ in Eq.\eqref{metri}, the metric $G_e$ is nonsingular on the boundary $V(q)=E$;
moreover, by varying the total energy $E$ we get a family of different metrics $g_J$, whereas
by choosing a convenient value of $\eta$, at different values of the energy
the metric $G_e$ remains the same. The consequence is that a comparison
among the geometries of the submanifolds of $(M\times\mathbb{R}, G_e)$---where the
geodesic flows of different energies ``live''---is meaningful. To the contrary, this
is not true with $(M_E,g_J)$. In some cases, the possibility of making this kind of
comparison can be important.
With respect to the Eisenhart metric $g_e$ on $M\times \mathbb{R}^2$ in the previous section,
the metric $G_e$ on $M\times \mathbb{R}$ defines a somewhat richer geometry, for
example the scalar curvature of $g_e$ is identically vanishing, which is not
the case of $G_e$.

In the case of a diagonal kinetic-energy metric, i.e. $a_{ij}\equiv\delta_{ij}$,
the only non vanishing Christoffel symbols are
\begin{equation}
\Gamma_{00}^i=\frac{(\partial V/\partial q^i)}{[2V(q)+ 2\eta]^2},~~~~~
\Gamma_{i0}^0= - \frac{(\partial V/\partial q^i)}{[2V(q)+ 2\eta]}\ ,
\end{equation}
whence the geodesic equations
\begin{eqnarray}
\frac{d^2q^0}{ds^2} +\Gamma^0_{i0}
\frac{dq^i}{ds}\frac{dq^0}{ds} +\Gamma^0_{0i}
\frac{dq^0}{ds}\frac{dq^i}{ds}&=&  0\ ,
\label{Ieqgeo0}\\
\frac{d^2q^i}{ds^2} +\Gamma^i_{00}
\frac{dq^0}{ds}\frac{dq^0}{ds} & =&  0\ ,
\label{Ieqgeoi}
\end{eqnarray}
which, using the affine parametrization of the arc length with time, i.e.,
$ds^2=2(E+\eta) dt^2$, with $(dq^0/dt)=2[V(q)+\eta]$ from (\ref{q(N+1)}), give
\begin{eqnarray}
\frac{d^2q^{0}}{dt^2} & = &  2 \frac{d{ V}}{dt}\ , \nonumber\\
\frac{d^2q^i}{dt^2} & = & - \frac{\partial V}{\partial q_i},~~~~~~i=1,\dots,N
~,\label{Ieqgeoit}
\end{eqnarray}
respectively.  The first equation is the differential version of (\ref{q(N+1)}), and
equations (\ref{Ieqgeoit}) are Newton's equations of motion.

\subsubsection{Curvature of \texorpdfstring{$(M\times\mathbb{R},G_e)$}{MxR,Ge}}
The basic curvature properties of the Eisenhart metric $G_e$
can be derived by means of the Riemann curvature tensor, which is found to have
the non-vanishing components 
\begin{equation}
R_{0i0j} = \frac{\partial_i\partial_j V}{(2V + 2\eta)^2} -
\frac{3(\partial_iV)(\partial_j V)}{(2V + 2\eta)^3}\ ,
\label{riemEisenG}
\end{equation}
whence, after contraction, using
$G^{00}=2V+2\eta$ the components of the Ricci tensor are 
found to be
\begin{eqnarray}
R_{kj}&=&\frac{\partial_k\partial_j V}{(2V + 2\eta)} -
\frac{3(\partial_kV)(\partial_j V)}{(2V + 2\eta)^2}\ ,\nonumber\\
R_{00}&=&\frac{\triangle V}{(2V + 2\eta)^2} -
\frac{3\Vert\nabla V\Vert^2}{(2V + 2\eta)^3}\ ,
\label{ricciEisenG}
\end{eqnarray}
where $\triangle V = \sum_{i=1}^N {\partial^2 V}/{\partial q^{i\, 2}}$, and thus
we find that the Ricci curvature at the point $q\in M\times\mathbb{R}$ and
in the direction of the velocity 
vector $\dot q$ is
\begin{equation}
K_R(q, \dot q)=\triangle V + R_{ij}\dot q^i \dot q^j
\end{equation}
and the scalar curvature at $q\in M\times\mathbb{R}$ 
is
\begin{equation}
{\mathcal R}(q) =\frac{\triangle V}{(2V + 2\eta)} -
\frac{3\Vert\nabla V\Vert^2}{(2V + 2\eta)^2}\ .
\label{scalarGe}
\end{equation}

\subsubsection{Geodesic Spread Equation for the Eisenhart Metric \texorpdfstring{$G_e$}{Ge}}
Let us now give the explicit form of Eq.(\ref{geod-dev}) in the case of
$(M\times\mathbb{R},G_e)$, the enlarged configuration space-time equipped
with one of the Eisenhart metrics. Since the nonvanishing Christoffel
coefficients are $\Gamma^i_{00}$ and $\Gamma^0_{0i}$, then using the affine
parametrization of the arc length with physical time, we obtain
\begin{eqnarray}\label{jlcge}
\frac{d^2J^k}{dt^2} +\frac{2(\partial_k V)}{2V+2\eta}\frac{dJ^0}{dt} +
\left[\partial_{kj}^2V -\frac{4(\partial_kV)(\partial_jV)}{2V+2\eta}\right] J^j &=& 0\ ,\nonumber
\\
&&\\
\frac{d^2J^0}{dt^2} -\frac{2(\partial_i V)\dot q^i}{2V+2\eta}\frac{dJ^0}{dt} -
2(\partial_iV)\frac{dJ^i}{dt} -
\left[\partial_{ij}^2V -\frac{2(\partial_iV)(\partial_jV)}{2V+2\eta}\right]\dot q^i J^j &=& 0\ ,\nonumber
\end{eqnarray}
where the indexes $i,j,k$ run from $1$ to $N$. These equations have not
yet been used to tackle Hamiltonian chaos, but are certainly worth to be investigated.

As reported in  Ref.\cite{cerruti1997lyapunov}, the JLC equation in Eq.\eqref{jlce} is rather complicated for the kinetic energy (Jacobi) metric in \eqref{metri},  it considerably simplifies to \eqref{dyntanT} for $(M\times{\mathbb R}^2,g_e)$, and displays an intermediate level of complexity for $(M\times \mathbb{R},G_e)$ as shown by Eqs.\eqref{jlcge}. This is related with a different degree of "richness" of the geometrical properties of the respective manifolds. It is therefore important to check whether all these geometrical frameworks provide the same information about regular and chaotic motions \cite{rick,cerruti1996geometric,cerruti1997lyapunov}, a necessary condition which a-priori could be questioned as it was done in Ref.\cite{cuervo2015non} even though the claims of this work have been proved wrong in \cite{loris}.

\section{Order and chaos in a paradigmatic two-degrees of freedom model with \texorpdfstring{$(M\times\mathbb{R},G_{e})$}{MxR,Ge}}
\begin{figure}[H]
	\includegraphics[height=8cm, width=8cm,keepaspectratio]{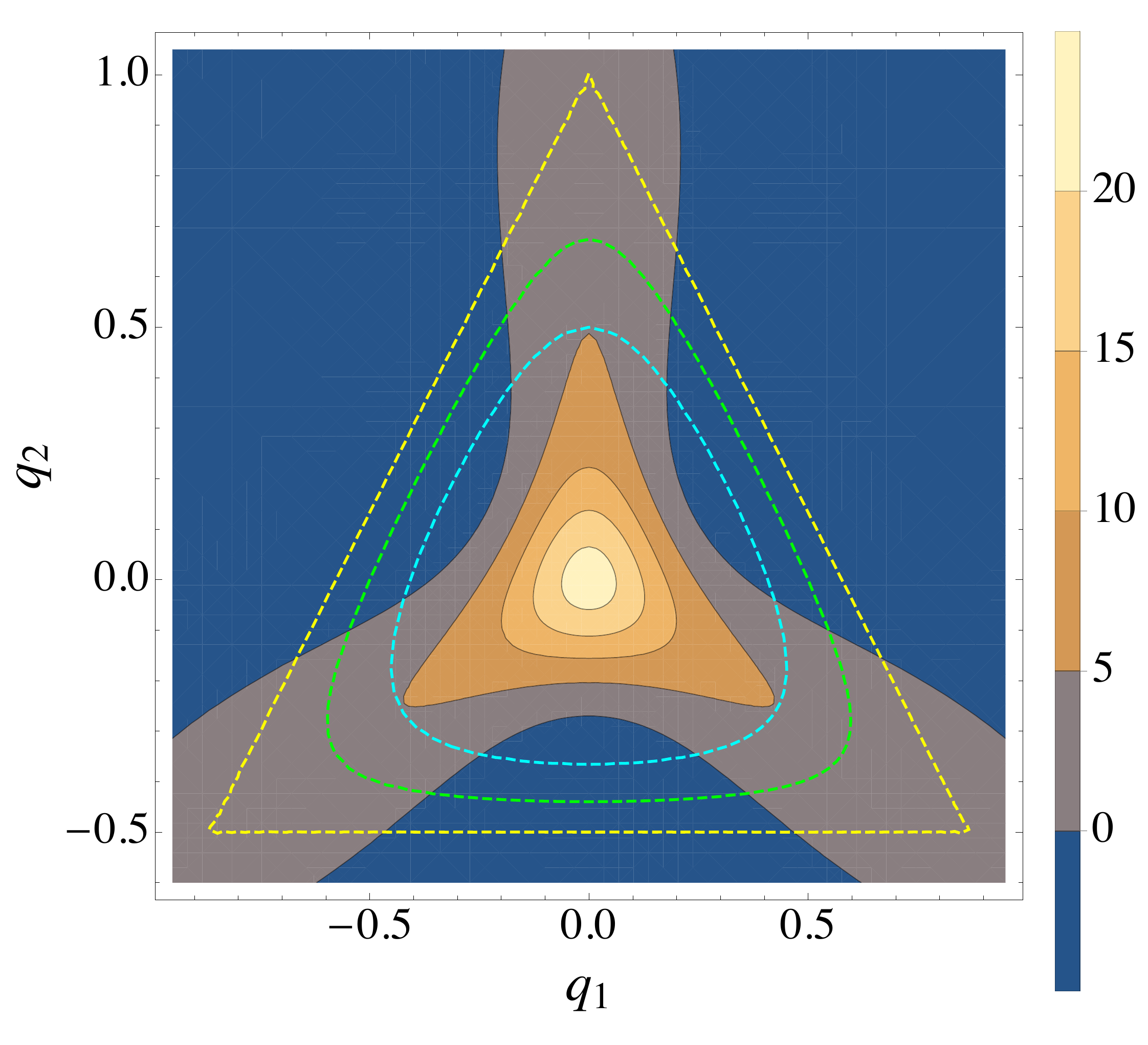}
		\includegraphics[height=8cm, width=8cm,keepaspectratio]{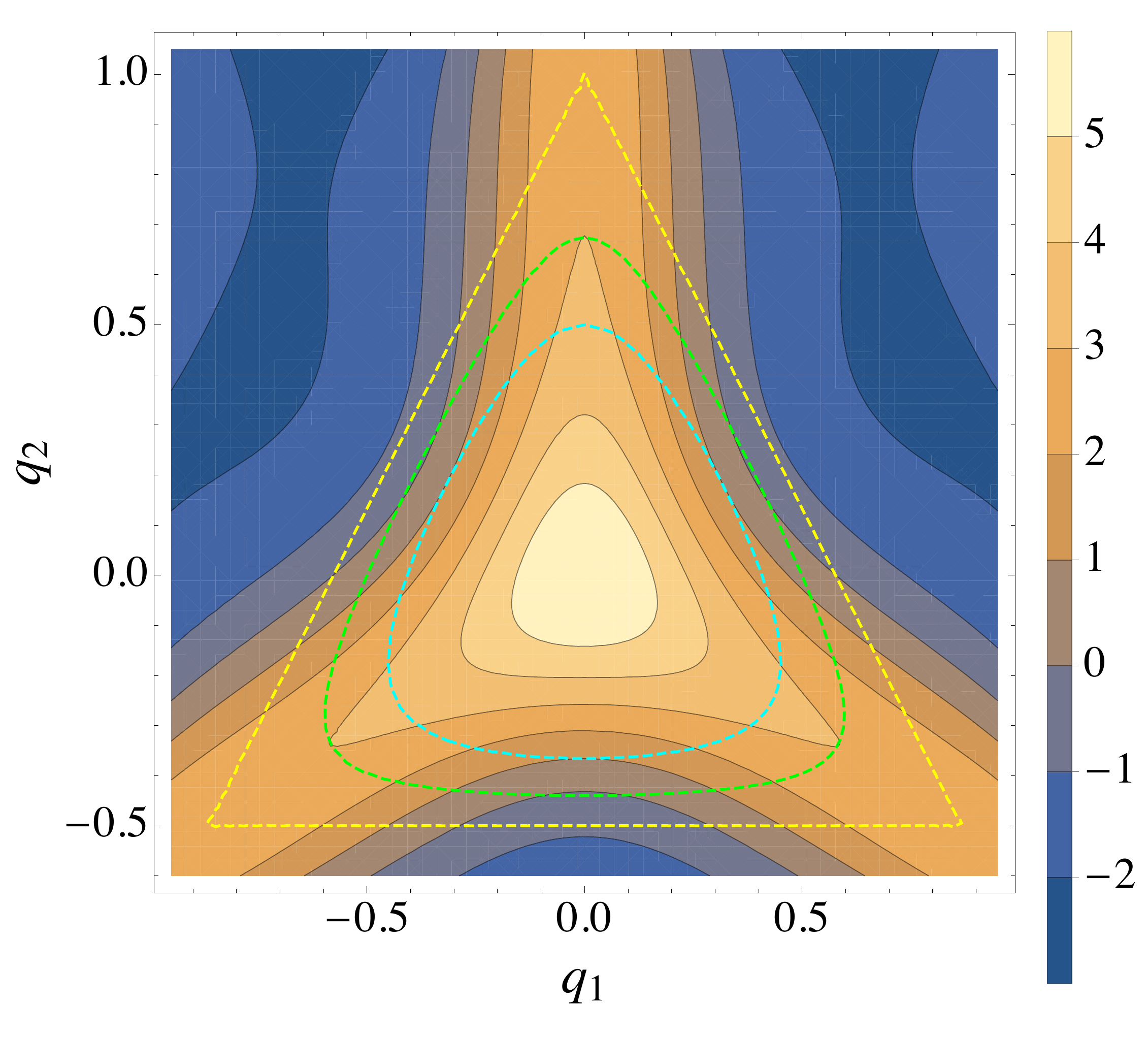}
	\caption{Configuration space of the H\'enon-Heiles model. The dashed lines represent the equipotential boundaries:  $V(q_1,q_2)=0.0833$ (cyan);  $V(q_1,q_2) = 0.125$ (green); $V(q_1,q_2) =0.1667$ (yellow). 
Left panel: $\eta = 0.045$. Right panel: $\eta = 0.1667$. The scale of colours represents different intervals of values of the scalar curvature given in Eq.\protect\eqref{scalarGe}.}
	\label{henonFig1}
\end{figure}

\begin{figure}[H]
	\includegraphics[height=8cm, width=8cm,keepaspectratio]{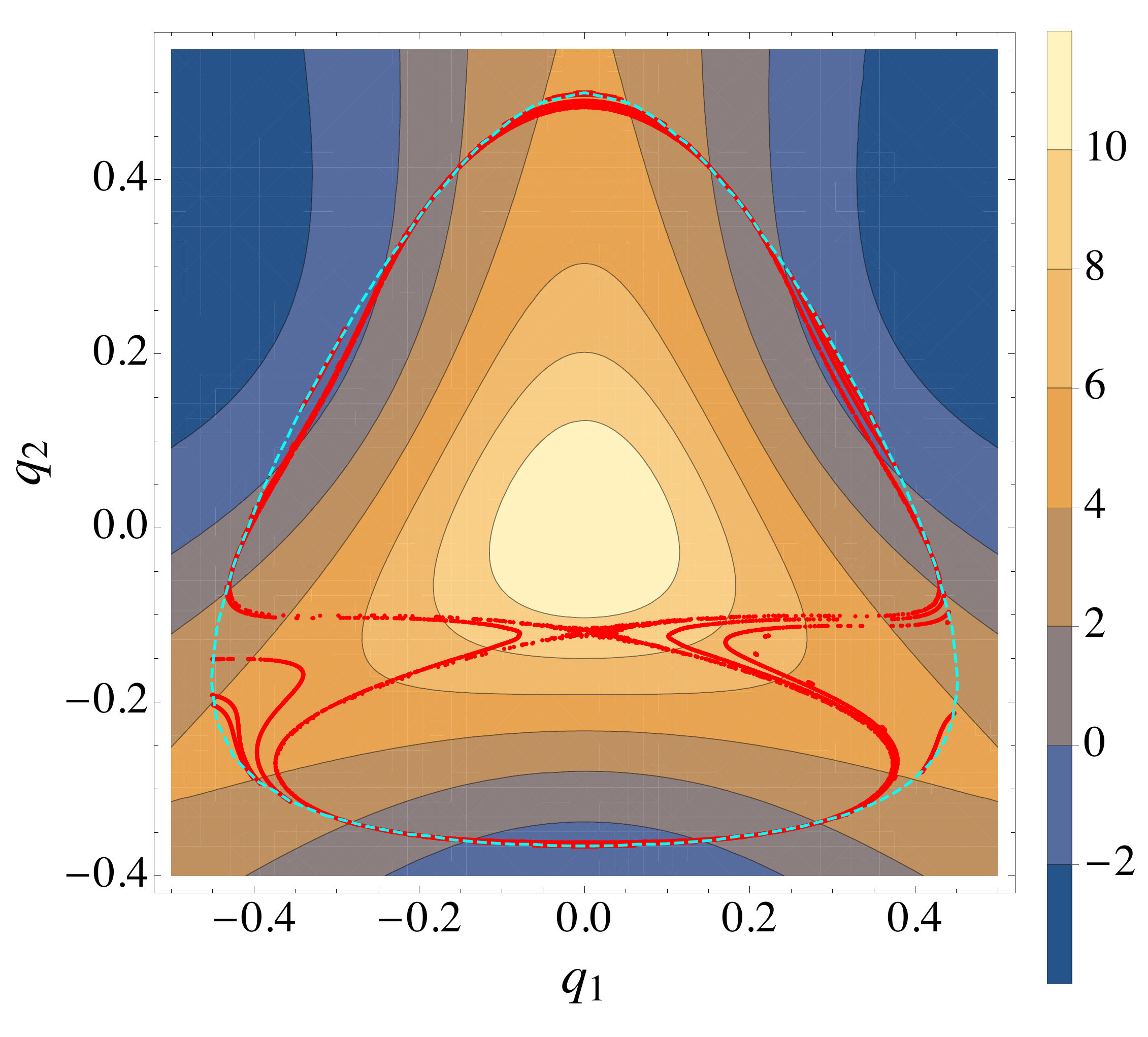}
		\includegraphics[height=8cm, width=8cm,keepaspectratio]{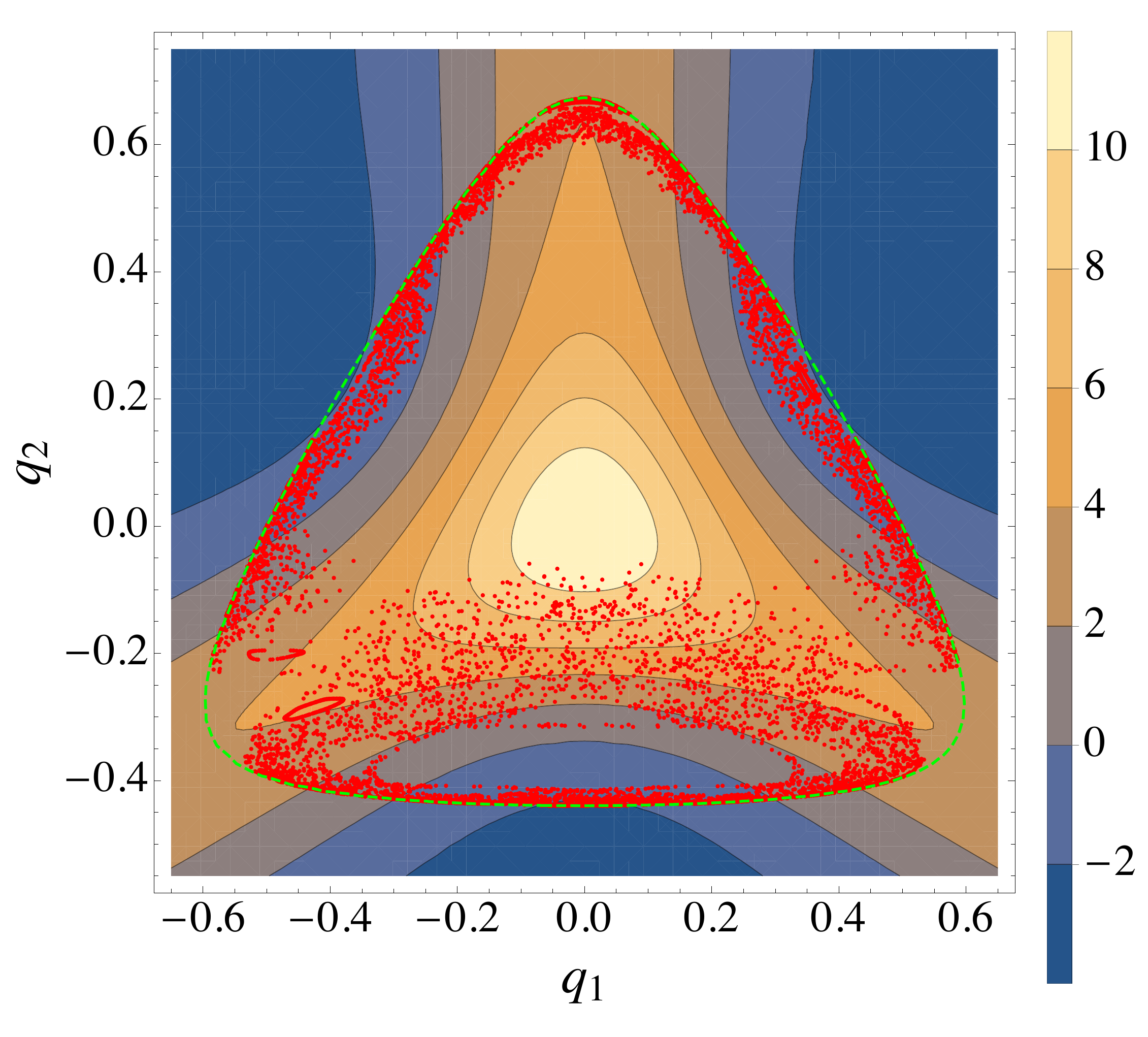}
		\centering
		\includegraphics[height=8cm, width=8cm,keepaspectratio]{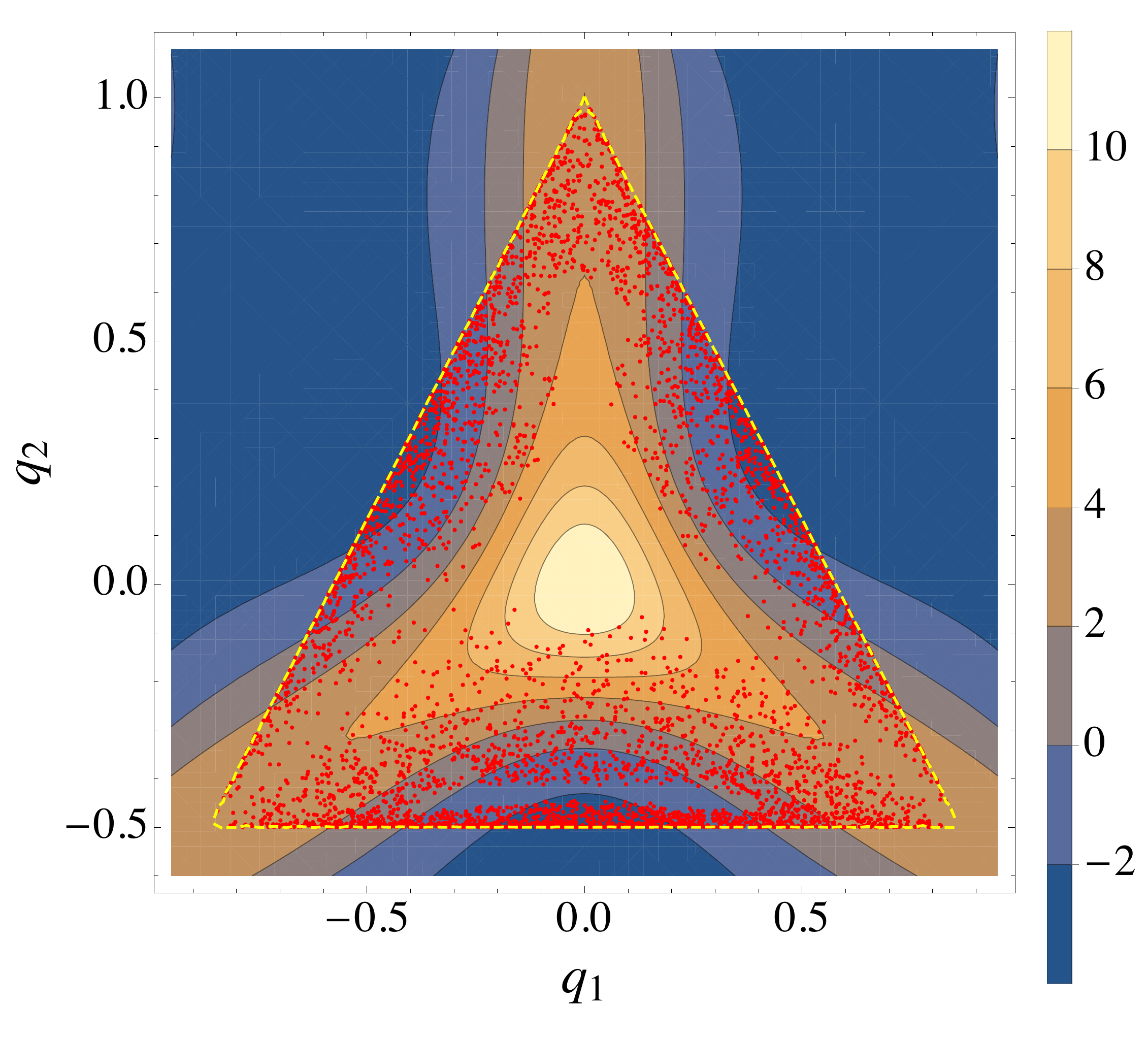}
	\caption{Superposition of the configuration space of the H\'enon-Heiles model with the surfaces of section of phase space trajectories. Red dots correspond to the crossing of the $(q_1,q_2)$ plane when $p_2=0$ and $p_1>0$. Upper left panel corresponds to $E= 0.0833$; upper right panel corresponds to $E = 0.125$; lower panel corresponds to $E=0.1667$. For all these cases $\eta =0.0833$.}
	\label{henonFig2}
\end{figure}

The first benchmarking is performed for a two-degrees of freedom system. In this case a paradigmatic candidate is the H\'enon-Heiles model described by the Hamiltonian 
\begin{equation}
{ H} = \frac{1}{2} \left( p_x^2 + p_y^2 \right) +
\frac{1}{2} \left( q_1^2 + q_2^2 \right) + q_1^2 q_2 - \frac{1}{3} q_2^3 \ .
\label{Henon-Heiles}
\end{equation}
In this case, the JLC equation for the Jacobi metric is exactly written in the form 
\begin{eqnarray}
\frac{d^2 J^\perp}{ds^2} &+ &\frac{1}{2}\left[ \frac{\triangle V}{(E - V)^2} + 
\frac{\Vert\nabla V\Vert^2}{(E - V)^3}\right] \, J = 0~,\label{eq_jacobi_2}\\
\frac{d^2 J^\parallel}{ds^2} &= &0
\end{eqnarray}
where the expression in square brackets is the scalar curvature of the manifold $(M_E,g_J)$, $g_J$ is the metric tensor whose components are in Eq.\eqref{metri}, $J^\perp$ and $J^\parallel$ are the components of the geodesic separation vector transversal and parallel to the velocity vector along the reference geodesic, respectively. It is well evident that this scalar curvature is always positive and that chaotic motions can only be the consequence of parametric instability due to the variability of the scalar curvature along the geodesics.
At first sight, the scalar curvature of $(M\times{\mathbb R},G_{e})$ given in Eq.\eqref{scalarGe} can take also negative values as is shown in Figure \ref{henonFig1}. On the one side this could add another source of dynamical instability to parametric instability, but, on the other side, the extension of regions of negative curvature depends on the value of the arbitrary parameter $\eta$ that enters the metric $G_e$, extension that can be arbitrarily reduced making its contribution to degree of chaoticity not intrinsic.
In Figure \ref{henonFig2} the plane $(q_1,q_2)$  is taken as surface of section of phase space trajectories when $p_2=0$ and $p_1>0$. 

At the lowest energy, $E= 0.0833$, when all the motions are regular, the trajectories are found to visit also regions of negative curvature, whereas at higher energies, $E = 0.125$ and $E=0.1667$, the chaotic trajectories considered display a large number of intersections in regions of positive curvature. In other words, the role of negatively curved regions does not appear
to play a relevant role in determining the chaotic instability of the dynamics.

\begin{figure}[H]
\centering
	\includegraphics[height=8.5cm, width=8cm,keepaspectratio]{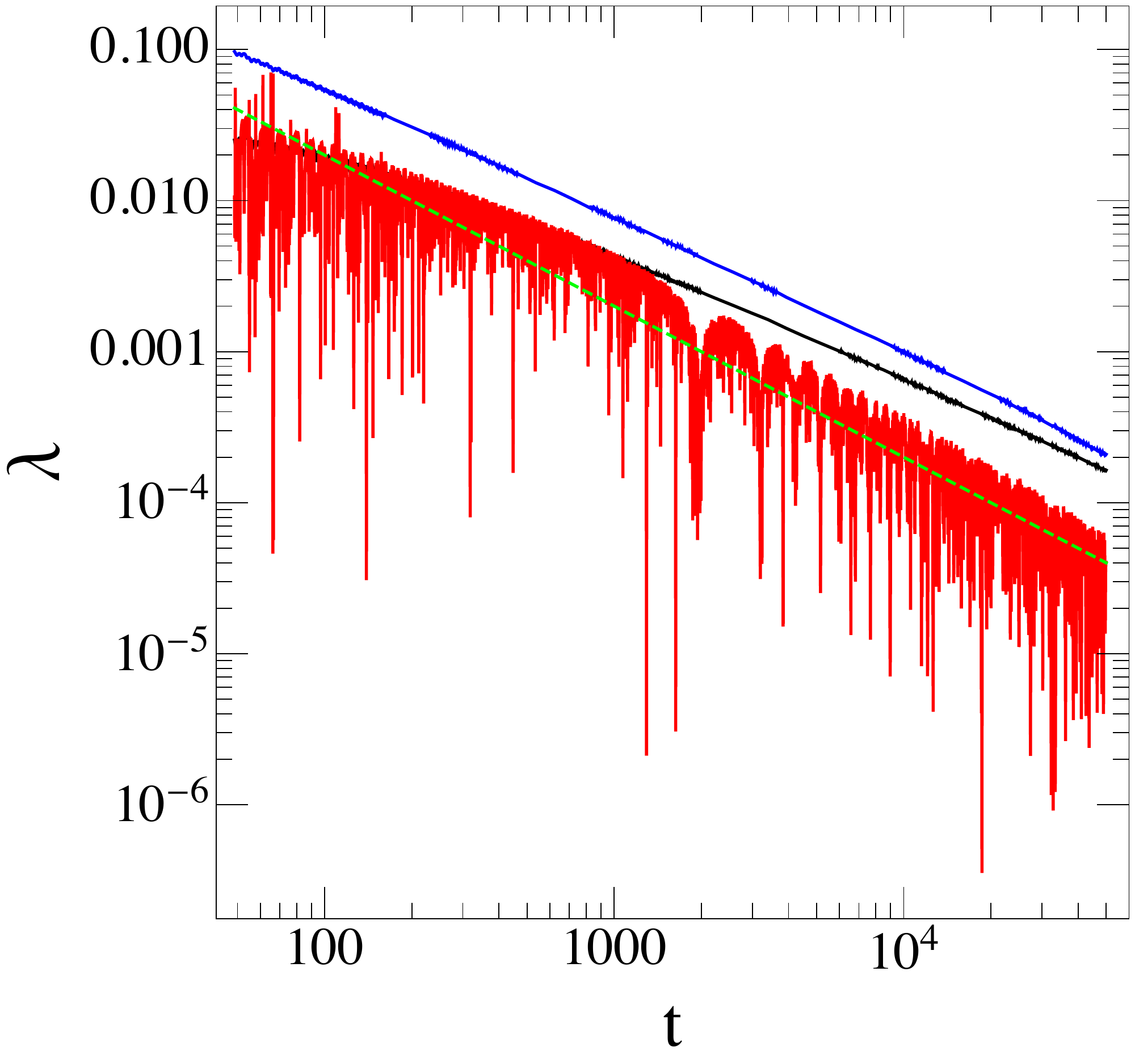}
	\includegraphics[height=8.5cm, width=8.3cm,keepaspectratio]{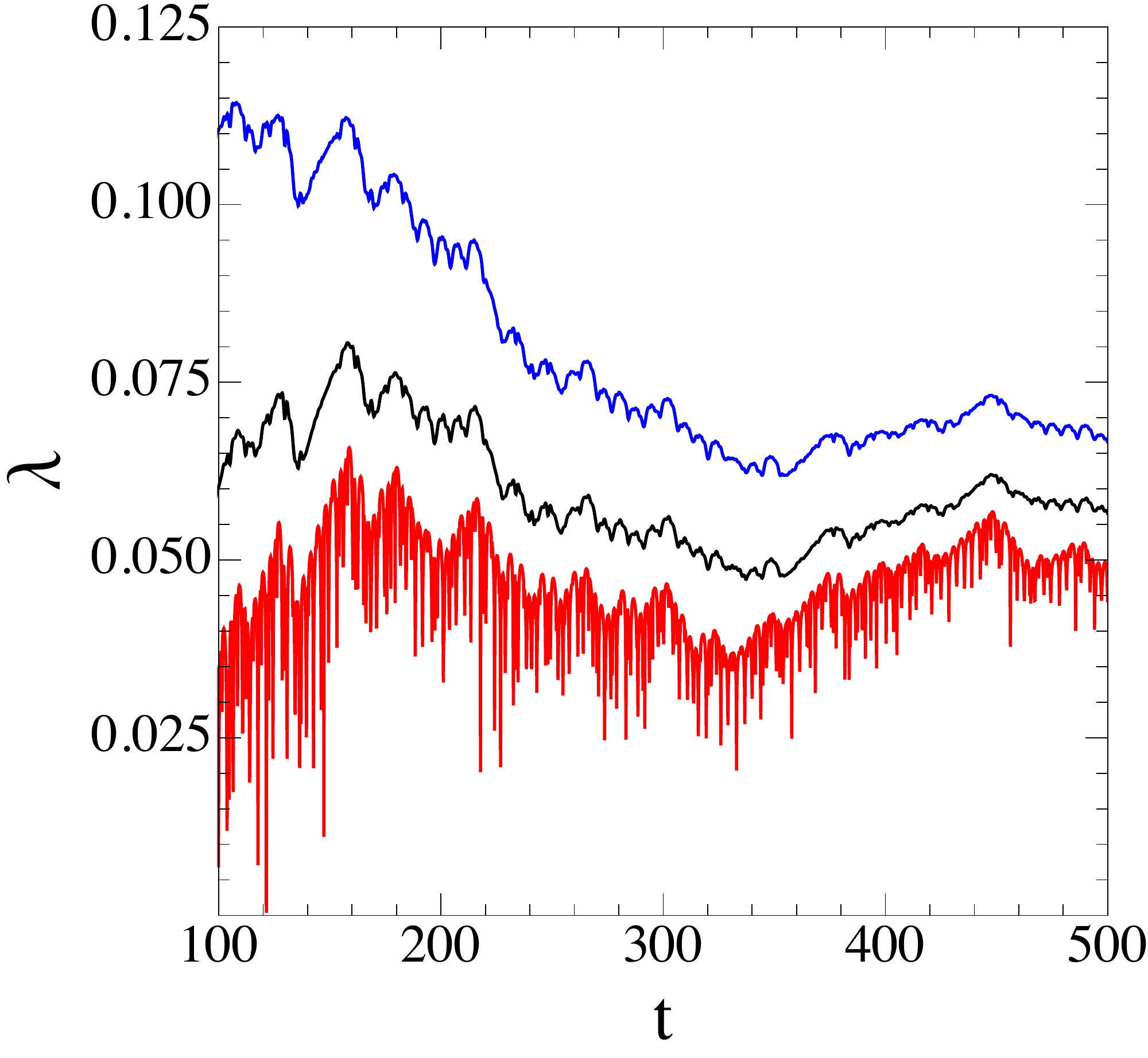}
	\caption{Numerical solutions of the tangent dynamics equation \protect\eqref{dyntanT} (black line) compared to the solution of 
	equation \protect\eqref{jlcge} (blue line), and to the solution of equation \protect\eqref{eq_jacobi_2} (red line). Left panel: $E=0.0833$, $\eta = 0.0833$ and the initial condition is point ($a$) of Figure 1 of \cite{cerruti1996geometric}. The dashed green line is the reference $t^{-1}$ slope for regular motions. Right panel: $E=0.125$, $\eta = 0.0833$ and the initial condition is point ($d$) of Figure 3 of \cite{cerruti1996geometric}. }
	\label{figure3}
\end{figure}

\begin{figure}[H]
	\includegraphics[height=8cm, width=8cm,keepaspectratio]{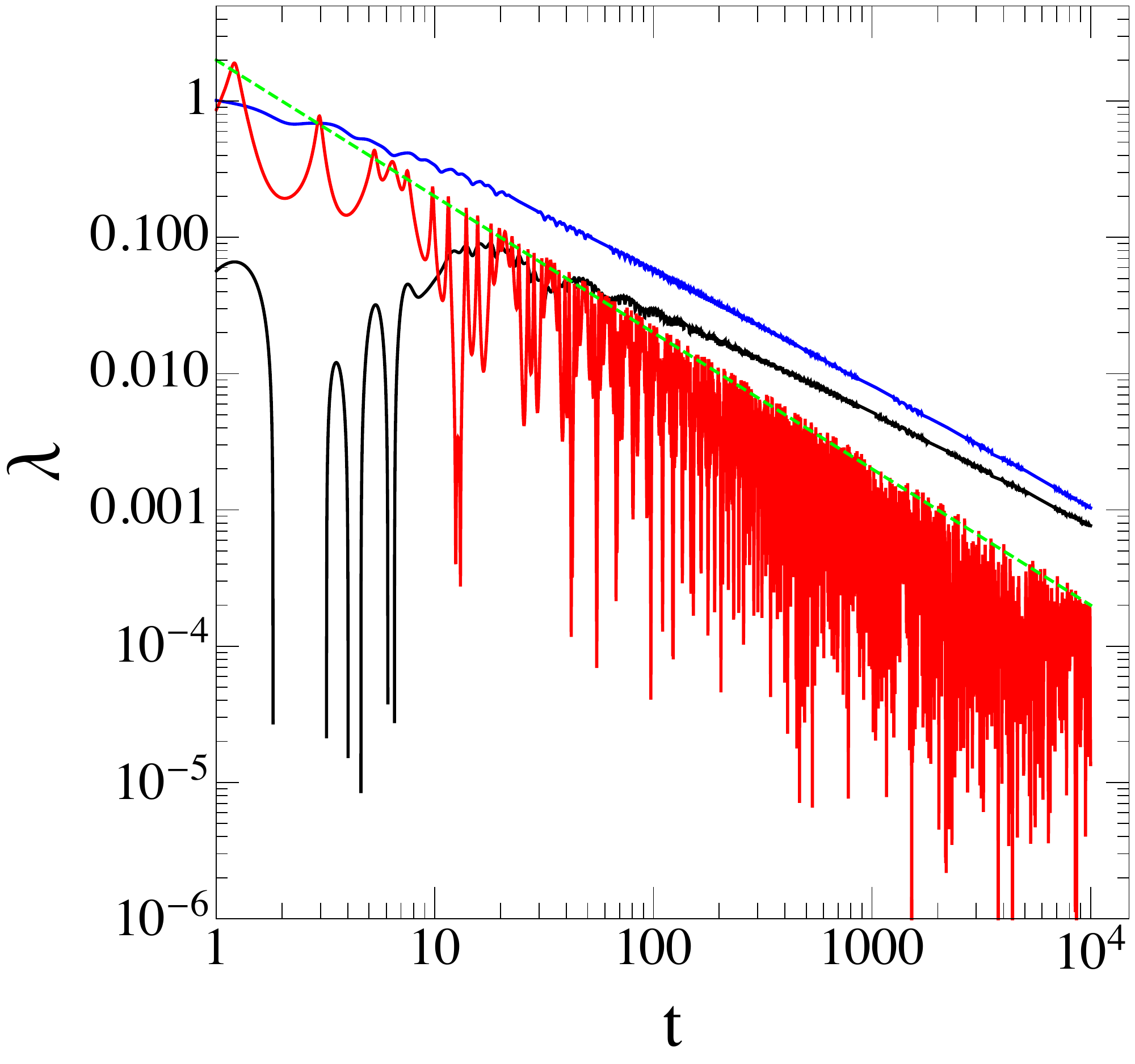}
		\includegraphics[height=7.9cm, width=7.9cm,keepaspectratio]{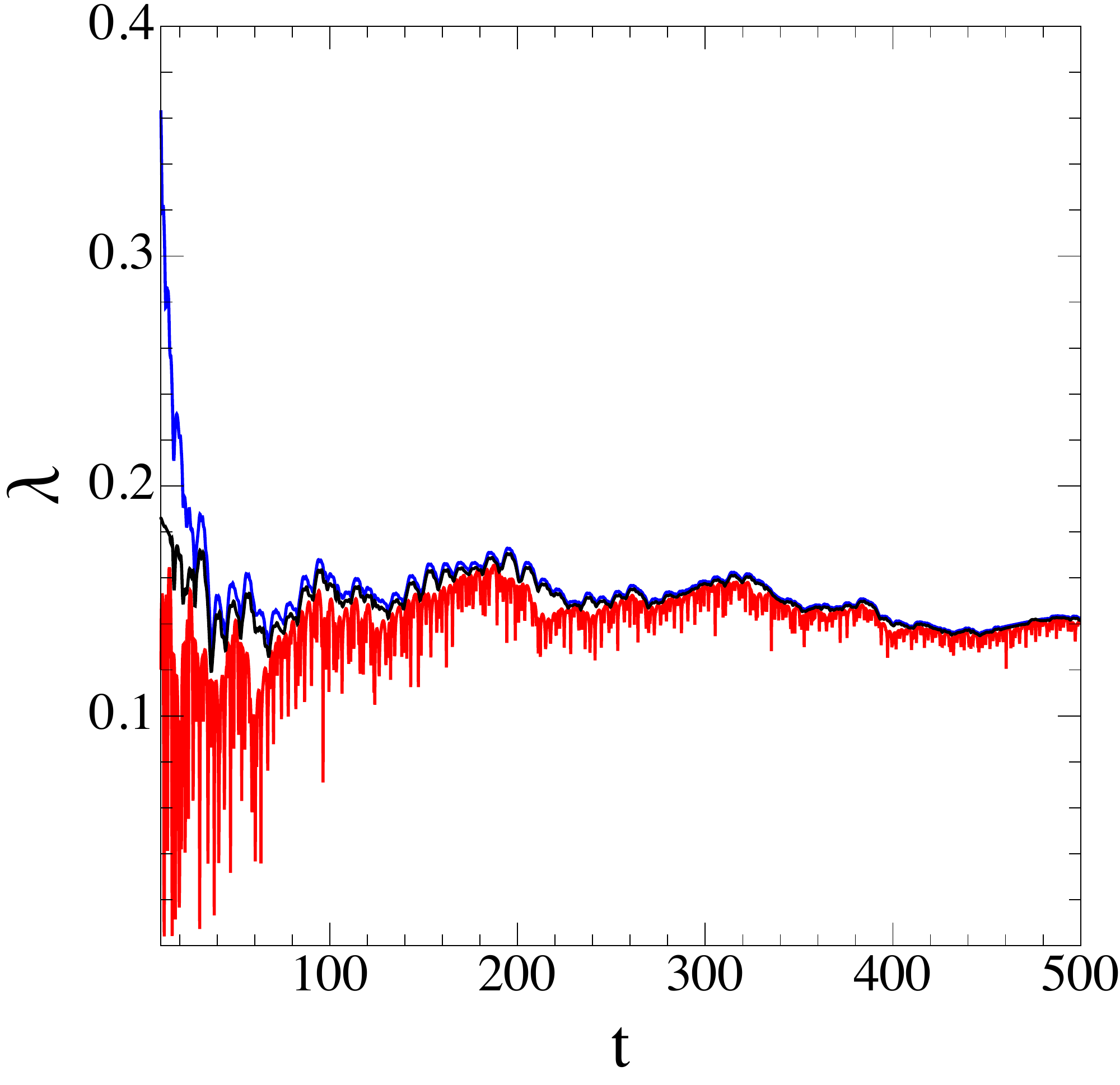}
	\caption{Numerical solutions of the tangent dynamics equation \protect\eqref{dyntanT} (black line) compared to the solution of 
	equation \protect\eqref{jlcge} (blue line), and to the solution of equation \protect\eqref{eq_jacobi_2} (red line). Here $E=0.1667$, $\eta = 0.0833$ and the initial condition for the left panel is point ($a$) of Figure 5 of \cite{cerruti1996geometric}, and for the right panel point ($c_2$) of the same Figure.}
	\label{figure4}
\end{figure}

As a matter of fact, the comparison of the results obtained by numerically integrating the stability equations \eqref{dyntanT}, \eqref{jlcge}, and \eqref{eq_jacobi_2} along with the equations of motion of the H\'enon-Heiles model, at different energies and initial conditions, show an excellent qualitative and quantitative agreement. The integration of the Hamilton equations of motion is performed with a symplectic integrator. The stability equations have been integrated with a fourth-order Runge-Kutta scheme. The choice of the energy values follows the historical paper by H\'enon-Heiles, and the initial conditions for regular and chaotic motions are chosen according to the selections in Ref.\cite{cerruti1996geometric}.
The quantity reported in Figures \ref{figure3} and \ref{figure4} is
\begin{equation}
\lambda (t) = \frac{1}{t}\log\left[\frac{\|{\dot J}(t)\|^2+\|J(t)\|^2}{\|{\dot J}(0\|^2+\|J(0)\|^2}\right]
\label{lambdaDT_XYmodel}
\end{equation}
where the separation vector $J$ is in turn the solution of the three different stability equations. 

The robustness of the results obtained by means of Eq.\eqref{jlcge} for the manifold $(M\times\mathbb{R},G_e)$ with respect to different choices of the free parameter $\eta$ has been checked and confirmed. It is in particular the close agreement between the results obtained with the Eqs.\eqref{jlcge} and \eqref{eq_jacobi_2} which confirms that chaos stems from parametric instability, because in the latter equation the scalar curvature is always positive. The right panel of Figure \ref{figure3} shows a clear qualitative agreement among the three patterns $\lambda(t)$ but some quantitative deviations that do not change neither with longer integrations not by changing the value of $\eta$ in the case of $\lambda(t)$ computed with \eqref{jlcge}. Perhaps such a discrepancy could stem from the inhomogeneity of the chaotic layer in phase space due to the presence of very small regular islands, inhomogeneity detected differently by the different JLC equations. Actually, this discrepancy is no longer observed at higher energy (right panel of Figure \ref{figure4}) when the chaotic layer seems more homogeneous. 
The reason why the geometrization of Hamiltonian dynamics by means of $(M\times \mathbb{R},G_e)$ can be of prospective interest relies on its intermediate geometrical "richness". 

On $(M\times{\mathbb R}^2,g_e)$ the scalar curvature is always vanishing, the Riemann curvature tensor is just the Hessian of the potential and the Ricci tensor has only one non-vanishing component, to the opposite, on $(M_E,g_J)$ the Riemann curvature tensor has ${\cal O}(N^4)$ non-vanishing components and at large $N$ the scalar curvature can happen to be overwhelmingly negative without affecting the degree of chaoticity of the dynamics. The geometry of  $(M\times\mathbb{R},G_e)$ is definitely richer than that of $(M\times{\mathbb R}^2,g_e)$ and less complicated than that of $(M_E,g_J)$, therefore, and mainly at large $N$, this framework can offer some computational advantage for more refined investigations about the geometric origin of parametric instability of the geodesics. Loosely speaking, to give an idea of what  a more refined geometrical investigation might mean, it has been shown \cite{book,cecmar} that integrability is related with the existence of Killing tensor fields on the mechanical manifolds, therefore the degree of breaking of the hidden symmetries associated with Killing tensor fields could be defined, investigated, and related with the existence of weak and strong chaos in Hamiltonian flows.

\section{One-dimensional \texorpdfstring{$XY$}{XY}-model in the Eisenhart metric \texorpdfstring{$(M\times\mathbb{R},G_{e})$}{MxR,Ge}}
Let us now proceed to investigate how Hamiltonian chaos is described in this geometric framework at a large number of degrees of freedom. This is shown for a specific model, the one-dimensional classical XY model. The reason for choosing this model is that it has a rich variety of dynamical behaviors: at low energy it is equivalent to a collection of weakly coupled harmonic oscillators, at asymptotically high energy it represents a set of freely rotating spins, at intermediate energies it displays a strongly chaotic dynamics, as witnessed by the whole spectrum of Lyapounov exponents \cite{JSP}. Moreover, for this model it was necessary to introduce an \textit{ad hoc} adjustment of an otherwise successful geometric-statistical model for the analytic computation of the largest Lyapounov exponent \cite{CasClePet} carried on in the framework $(M\times{\mathbb R}^2,g_e)$. It is thus interesting to check whether or not another geometric framework can allow to fix the problem more naturally.

The 1D $XY$ model, describes a linear chain
of $N$ spins/rotators constrained to rotate in a plane and coupled
by a nearest-neighbour interaction. This model is formally obtained by restricting
the classical Heisenberg model with $O(2)$ symmetry to one
spatial dimension.
The potential energy of the $O(2)$ Heisenberg model is
$V= -{\cal I}\sum_{\langle i,j \rangle}{\bf s}_{i}\cdot{\bf s}_{j}$, where the sum is
extended only over nearest-neighbour pairs, ${\cal I}$ is the
coupling constant, and
each ${\bf s}_{i}$ has unit modulus and rotates in the plane.
To each ``spin'' ${\bf s}_{i}= (\cos q_{i},\sin q_{i})$, the
velocity
${\bf \dot s}_{i}=(-\dot q_{i} \sin q_{i},
\dot q_{i} \cos q_{i})$ is associated, so that ${H}=\sum_{i=1}^{N}
\frac{1}{2}\dot{\bf s}_{i}^{2} - {\cal I} \sum_{\langle i,j \rangle}
{\bf s}_{i}\cdot {\bf s}_{j}$.
The Hamiltonian of this model is then
\begin{equation}
H(p,q) = \sum_{i=1}^{N}\frac{p_i^2}{2} + {\cal I}\sum_{i=1}^{N}[1-\cos(q_{i}-q_{i-1})]~,
\label{H_XY_Model}
\end{equation}
The canonical coordinates $q_i$ and $p_i$ are thus given the meaning of angular
coordinates and momenta. 
As already mentioned above, this Hamiltonian system has two integrable limits. In the
low-energy limit it represents a chain of harmonic
oscillators, as can be seen by expanding the potential energy in power
series 
\begin{equation}
{ H}(p,q)
\approx\sum_{i=1}^{N}\left[\frac{p_{i}^{2}}{2}+\frac{{\cal I}}{2}
(q_{i+1}-q_{i})^2  \right]~,
\label{hamrotE0}
\end{equation}
where $p_i = \dot q_i$, whereas in the high-energy limit it represents a system
of freely rotating objects, 
since the kinetic energy increases with total energy without bounds, at variance with  
potential energy which is bounded from above.

\subsection{Numerical solution of the JLC equation for \texorpdfstring{$(M\times\mathbb{R},G_{e})$}{MxR,Ge}}
\label{subsec_EisenhartMetricMxR}
\begin{figure}[H]
	\centering
	\includegraphics[height=10cm, width=10cm,keepaspectratio]{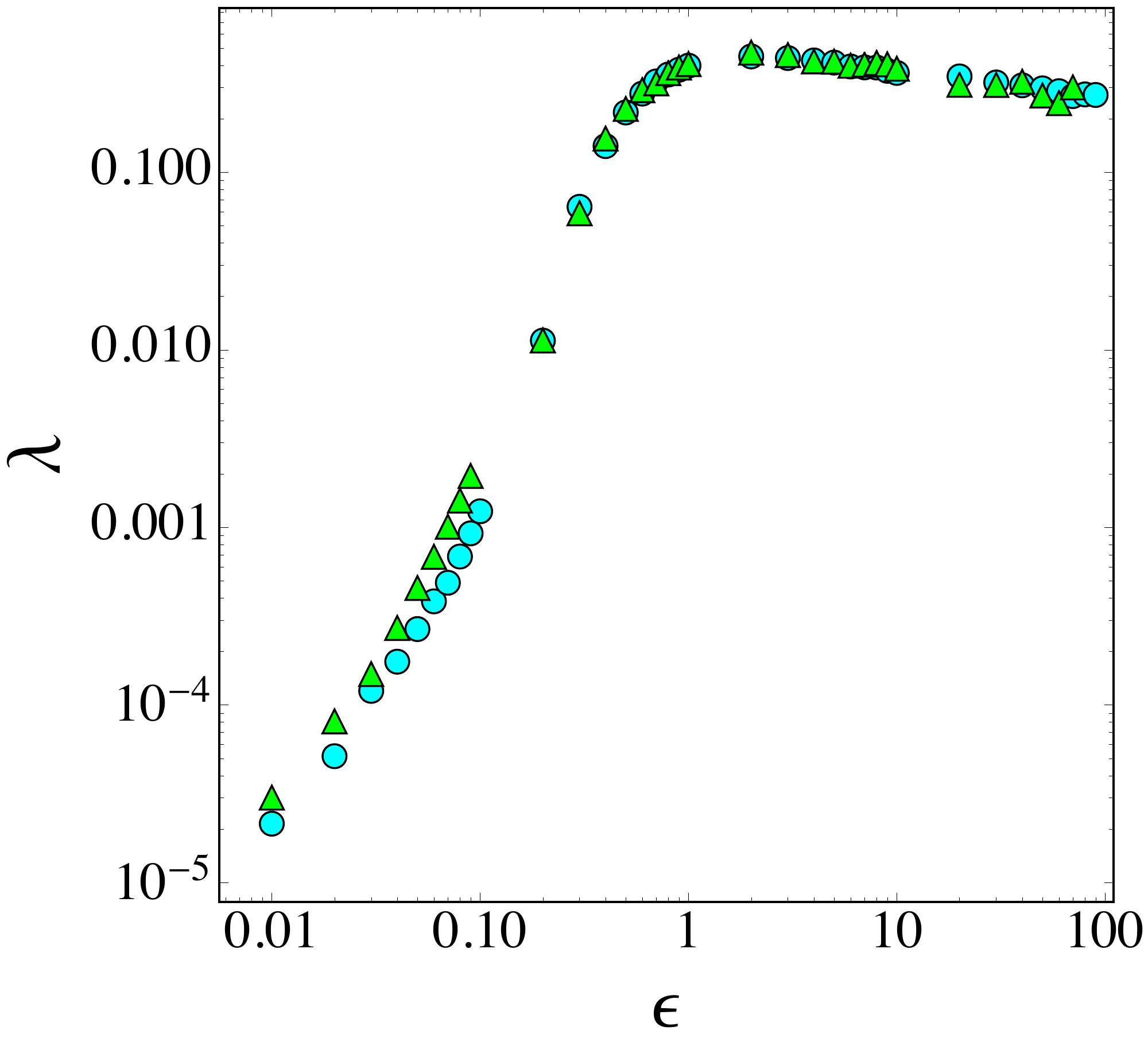}
	\caption{Lyapunov Exponents $\lambda_1$ (cyan circles) and $\lambda_{G}$ (green triangles) versus the  energy density 
	$\epsilon$ for a system of $N=150$ spins. The parameter $\eta$ has been set as $\eta=E$.}
	\label{comparison_dintan_eis1}
\end{figure}

Let us proceed by comparing the outcomes of the integration of the equations \eqref{dyntanT} and \eqref{jlcge} computed along the flow of the Hamiltonian \eqref{H_XY_Model}.
The standard tangent dynamics equations  \eqref{dyntanT} can be split as
\begin{eqnarray}\label{TanDyn}
\dot{J}^{i}_{q}&=&J^{i}_{p}\nonumber\\
\dot{J}^{i}_{p}&=&- Hess (V)_{ij} \ J^{j}_{q}
\end{eqnarray}
which explicitly read as 
\begin{eqnarray}
\dot{J}^{i}_{q}&=&J^{i}_{p}\label{tgDyn}\\
\dot{J}^{i}_{p}&=& - {\cal I}\cos(q_{i-1}-q_{i})J^{i-1}_{q}+{\cal I}[\cos(q_{i-1}-q_{i})+\cos(q_{i}-q_{i+1})]J^{i}_{q}-{\cal I}\cos(q_{i-1}-q_{i})J^{i+1}_{q}\ ,\nonumber
\end{eqnarray}
whence the Largest Lyapunov Exponent is worked out by computing 
\begin{equation}
\lambda_{1} =\lim_{t\rightarrow\infty} \frac{1}{t}\log\left[\frac{\|J_q (t)\|^2+\| J_p(t)\|^2}{\|J_q (0)\|^2+\| J_p(0)\|^2}\right] \ .
\label{lambdaDT_XYmodel_general}
\end{equation}
At the same time, the integration of the JLC equations \eqref{jlcge}, by setting ${J}=(J^{0},J^{i})$, 
and choosing $\eta=E$, yields another estimate of the instability exponent through the analogous definition 
\begin{equation}
\lambda_{G}=\lim_{t\rightarrow\infty} \frac{1}{t}\log\left[\frac{\|{J}(t)\|^{2}_{G_{e}}+\|\dot{{J}}(t)\|^{2}_{G_{e}}}{\|{J}(0)\|^{2}_{G_{e}}+\|\dot{{J}}(0)\|^{2}_{G_{e}}}\right] \ .
\label{lambda_G_XY_Model}
\end{equation}
We have solved the equations of motion of the 1D XY model (setting ${\cal I}=1$) and the tangent dynamics equations \eqref{tgDyn} by using a bi-lateral symplectic algorithm \cite{lapo}.  The JLC equations \eqref{jlcge} have been solved by using a third-order predictor-corrector algorithm. Periodic boundary conditions have been considered. Random initial conditions have been adopted by taking the $q_{i}$ randomly distributed in the interval $[0,2\pi ]$, and by taking the $p_{i}$ gaussian-distributed and suitably scaled so as to complement with the kinetic energy the difference between the total energy initially set and the initial value of the potential energy resulting from the random assignment of the $q_{i}$.
Figure \ref{comparison_dintan_eis1} shows the comparison between the results obtained at different values of the energy density 
$\epsilon=E/N$ for $\lambda_{1}(\epsilon)$ and $\lambda_{G}(\epsilon)$ defined above. 
It is well evident that the results so obtained are globally in very good agreement.  At energy densities in the interval between  $\epsilon\simeq 0.2$ and  $\epsilon\simeq100$ the agreement is perfect, whereas at lower energy densities, below $\epsilon\simeq 0.2$, small discrepancies are found which seem due to a slower time-relaxation of  $\lambda_{G}(t)$ with respect to 
$\lambda_{1}(t)$.

\begin{figure}[H]
	\centering
	\includegraphics[height=9cm, width=9cm,keepaspectratio]{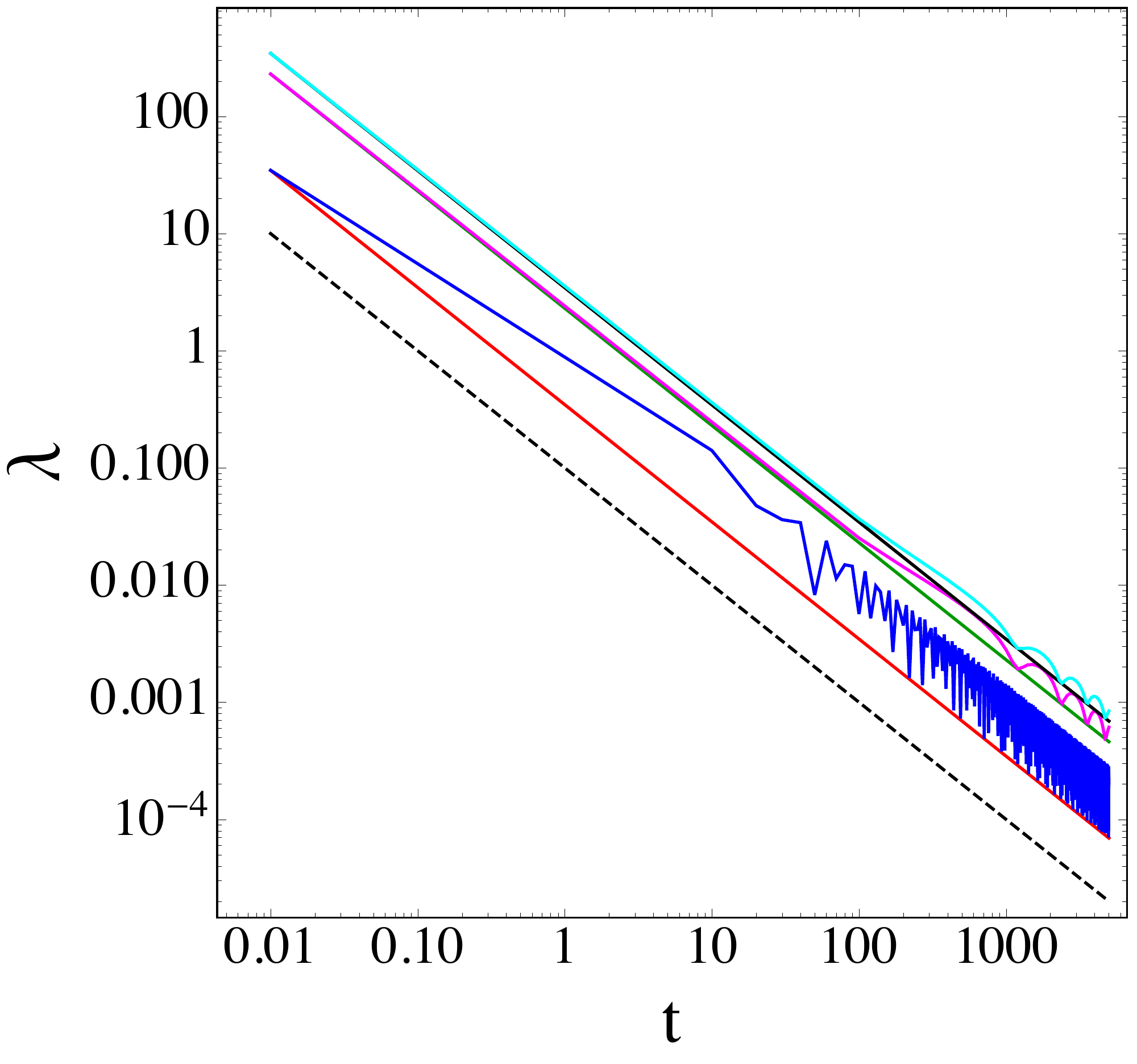}
	\caption{Lyapunov Exponents $\lambda_1(t)$ (red, green and black lines) versus $\lambda_{G}(t)$ (blue, magenta and cyan  lines)  for a system of $N=2,100,1000$ harmonic oscillators, respectively. The black dashed line is the $t^{-1}$ reference slope for a regular dynamics. Here $\epsilon = 1$ and $\eta=E$.}
	\label{LogLog_exp_lyap_oscillatori}
\end{figure}

Of course, an unavoidable check of consistency has to be performed on an integrable dynamics. This check has been performed
on the flow of the Hamiltonian \eqref{hamrotE0}. The results obtained with the equations \eqref{dyntanT} and \eqref{jlcge} are reported in Figure \eqref{LogLog_exp_lyap_oscillatori}. As expected for non-chaotic dynamics, it is found that $\lambda_{1}(t)$ decays as a straight line of slope $-1$ in double logarithmic scale, and $\lambda_{G}(t)$ decays with an oscillating pattern with a $t^{-1}$ envelope. This has been checked at different $N$ and energy values. Some cases are reported in Figure \ref{LogLog_exp_lyap_oscillatori}.

\section{The effective scalar model for the JLC equation}
In \cite{CasClePet} an effective scalar approximation of the JLC equation \eqref{jlce} has been worked out under some suitable hypothesis. In a nutshell, at large $N$ under an hypothesis of quasi-isotropy - meaning that a coarse-grained mechanical manifold appears as a constant curvature isotropic manifold - with broad spatial spectrum of curvature variations at a finer scale, the evolution of the norm of the geodesic separation vector is described by a stochastic oscillator equation 
\begin{equation*}
\frac{d^{2}\psi (s)}{ds^{2}}+\left[ \langle k_{R}\rangle +\langle\delta^{2}k_{R}\rangle^{1/2}\eta(s)\right] \psi(s)=0
\end{equation*}
where  $\eta(s)$ a $\delta$-correlated gaussian stochastic process of zero mean and unit variance, and 
\begin{equation*}
\begin{split}
\langle k_{R}\rangle &=\frac{1}{N-1}\langle K_{R}\rangle \\
\langle\delta^{2}k_{R}\rangle^{1/2}&=\frac{1}{N-1}(\langle K_{R}^{2}\rangle -\langle K_{R}\rangle^{2})
\label{curvature_fluctuation_eisenhart}
\end{split}
\end{equation*}
where $K_R$ is the Ricci curvature of the mechanical manifold under consideration, and the averages are meant along a reference geodesic or as microcanonical averages on suitable energy surface $\Sigma_E$. By putting $k_{0}=\langle k_{R}\rangle$, 
$\sigma=\langle\delta^{2}k_{R}\rangle^{1/2}$,  
\begin{equation}
\begin{split}
\tau_{1}&=\Big\langle\frac{dt}{ds}\Big\rangle\frac{\pi}{2\sqrt{k_{0}+\sigma}}\\
\tau_{2}&=\Big\langle\frac{dt}{ds}\Big\rangle\frac{k_{0}^{1/2}}{\sigma}
\label{characteristic_time_scale}
\end{split}
\end{equation}
and hence defining $\tau^{-1}=2(\tau_{1}^{-1}+\tau_{2}^{-1})$, an analytic expression for a geometric Largest Lyapunov Exponent is given by \cite{CasClePet}
\begin{eqnarray}
\lambda(k_0,\sigma,\tau)  & = & \frac{1}{2}
\left(\Lambda-\frac{4k_0}
{3 \Lambda}\right)\ ,\nonumber\\
\Lambda &  = &
\left(\sigma^2\tau+\sqrt{\left(\frac{4k_0}{3}
	\right)^3+\sigma^4\tau^2}\,\right)^{1/3}\ .
\label{lambdaeisenhart}
\end{eqnarray}
This can be applied to the geometrization on the manifold $(M\times{\mathbb R},G_{e})$ of Hamiltonian dynamics. In this case the Ricci curvature reads as 
\begin{equation}
\begin{split}
K_{R}(s)=\frac{1}{2(E+\eta)}\left(\Delta V-\frac{3\|\nabla V\|^2}{2V+2\eta}+\frac{\partial^2_{kj}V\, \dot{q}^j \dot{q}^k}{2V+2\eta}-\frac{3\partial_{j}V\, \dot{q}^j \partial_{k}V\dot{q}^k}{(2V+2\eta)^2}\right) \equiv \frac{K_{R}(t)}{2(E+\eta)}
\label{riccicurvatureeisenhart}
\end{split}
\end{equation}
and using the arc-length parametrization $ds^{2}=2(E+\eta)dt^{2}$ with physical time, we can compute by means of Eqs.\eqref{lambdaeisenhart} an analytic prediction of $\lambda_{G}(\epsilon)$ for $(M\times{\mathbb R},G_{e})$ and compare it to the outcome obtained  for  $(M\times{\mathbb R}^{2},g_{e})$.

\begin{figure}[H]
	\centering
	\includegraphics[height=9cm, width=9cm,keepaspectratio]{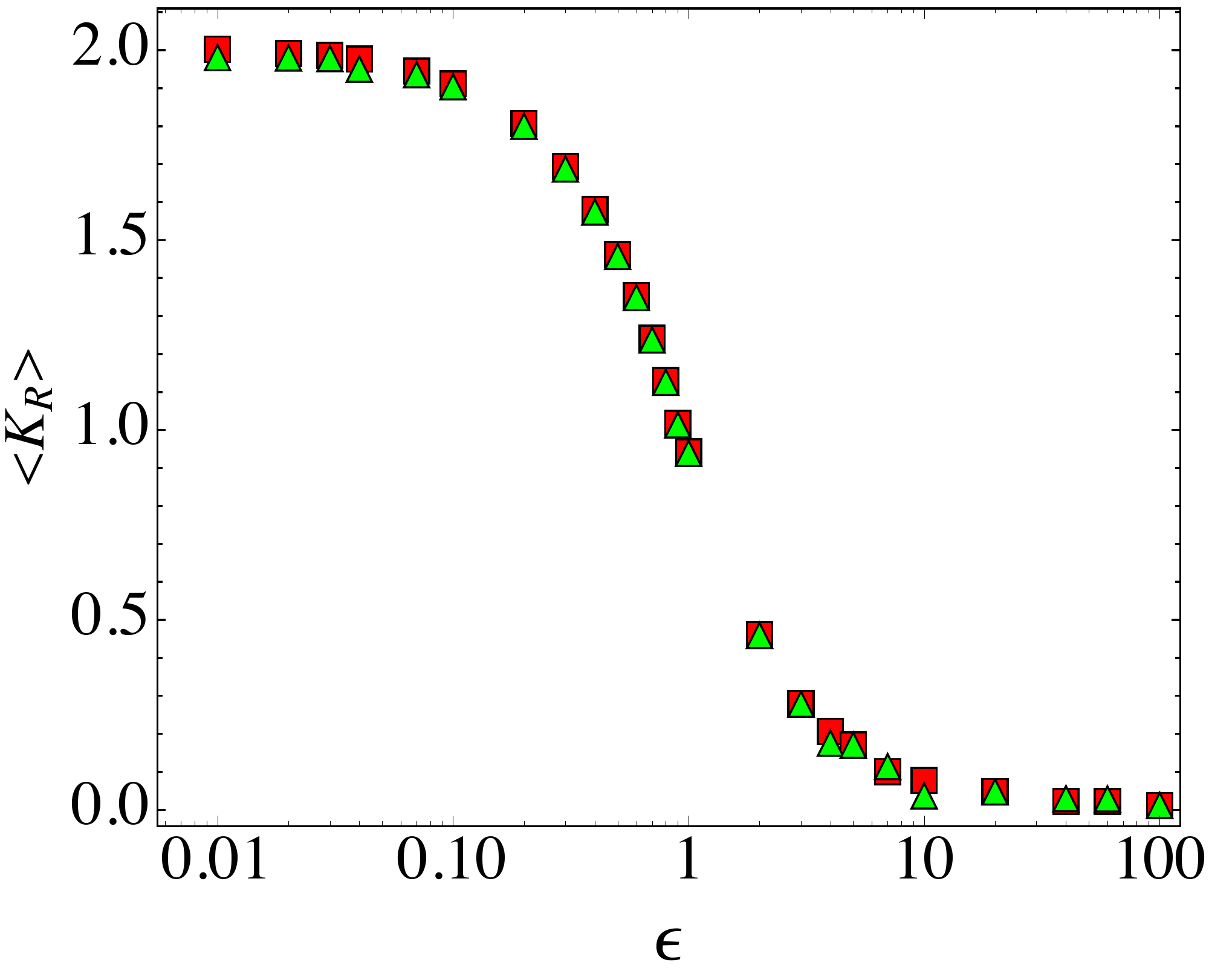}
	\caption{Average of Ricci curvature $\langle K_{R}\rangle$ of ${M\times{\mathbb R}^{2}}$ (red squares) and 
	of $M\times{\mathbb R}$ (green triangles), respectively, vs energy density $\epsilon$ for a system of $N=150$. 
	Here $\eta = E$.}
	\label{average_curvature_MxR_MxR2}
\end{figure}

\begin{figure}[H]
	\centering
	\includegraphics[height=9cm, width=9cm,keepaspectratio]{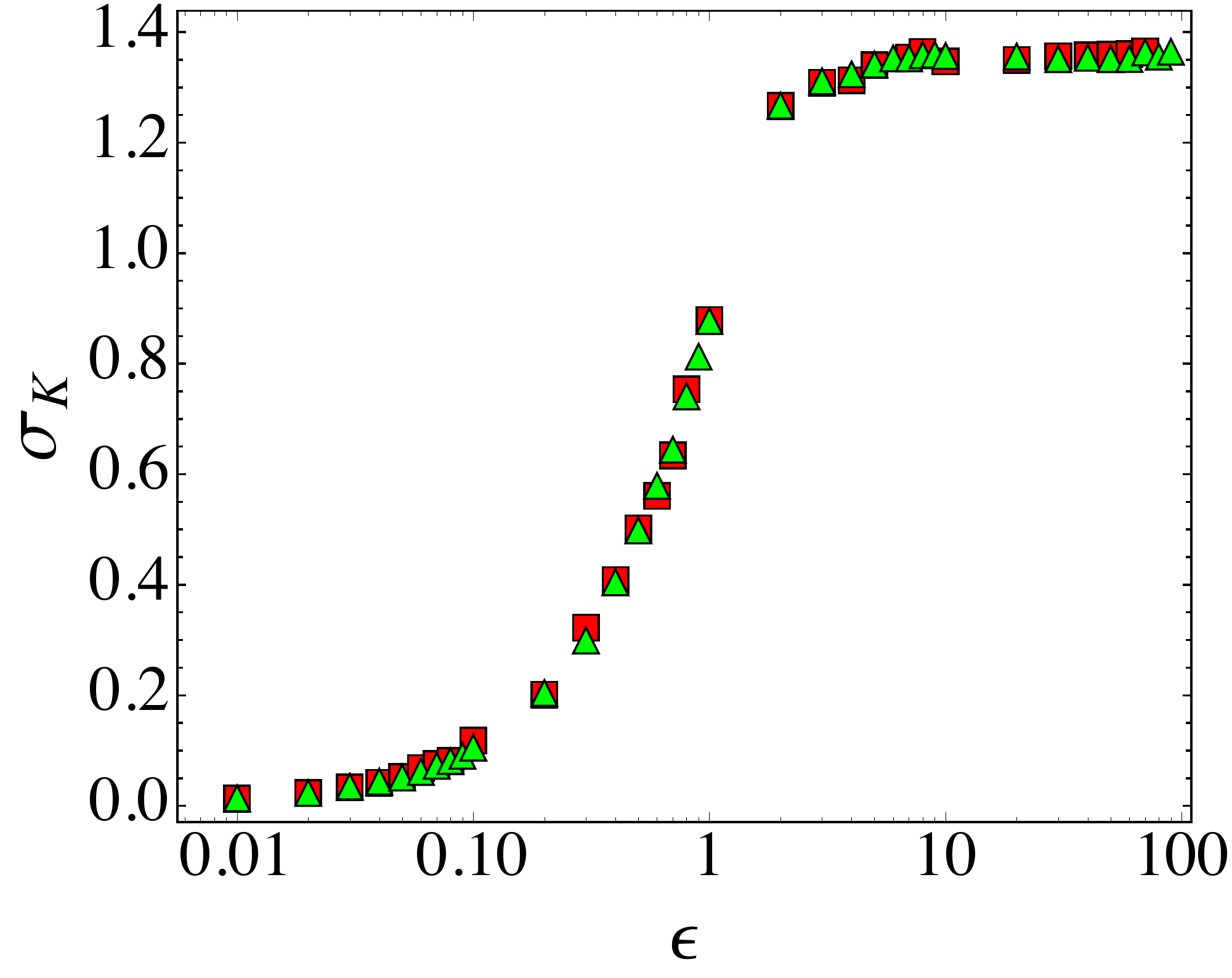}
	\caption{Average variance of the Ricci curvature $\sigma_K$ of ${M\times{\mathbb R}^{2}}$ (red squares) and 
	of $M\times{\mathbb R}$ (green triangles) vs energy density $\epsilon$ for a system of $N=150$ particles. 
	Here $\eta = E$.}
	\label{fluctuation_curvature_MxR_MxR2}
\end{figure}

\begin{figure}[H]
	\centering
	\includegraphics[height=9cm, width=9cm,keepaspectratio]{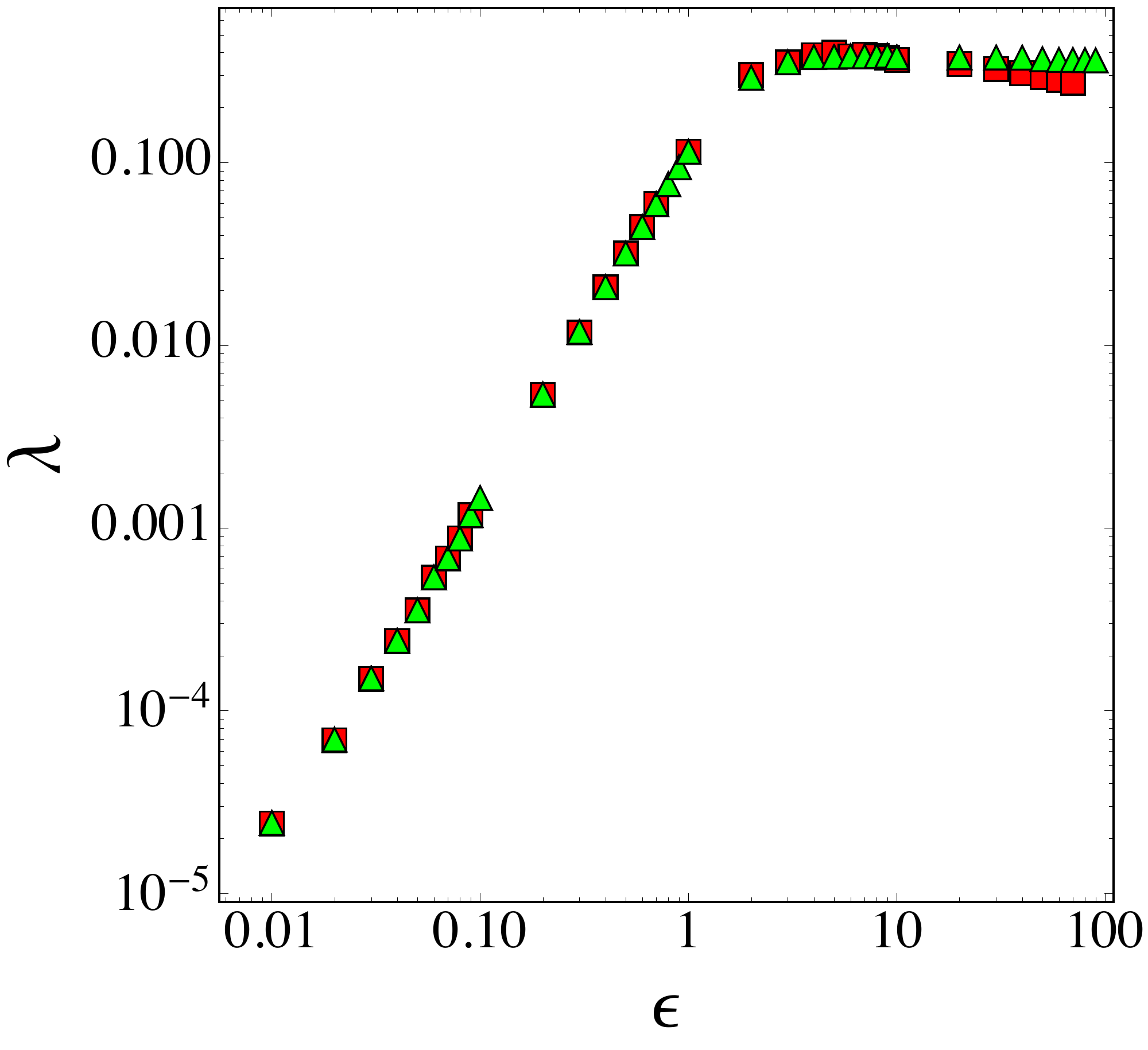}
	\caption{Geometric Lyapunov Exponents $\lambda$ $\lambda$ worked out for  
	 ${M\times{\mathbb R}^{2}}$ (red squares) and 
	for $M\times{\mathbb R}$ (green triangles) vs energy density $\epsilon$, for a system of $N=150$ particles. 
	Here $\eta = E$.}
	\label{comparison_geo_exp_MxR_MxR2}
\end{figure}

The first step consists in computing the average Ricci curvature and its variance of the two manifolds at different values of the energy density. We can limit these computations to one single choice of $N$ for which the asymptotic values of 
$\langle k_{R}\rangle$ and $\langle\delta^{2}k_{R}\rangle$ are already attained (see \cite{CasClePet}). Moreover, for non-integrable systems, after the Poincar\'e-Fermi theorem, all the constant energy surface is accessible to the dynamics, and since chaos entails phase space mixing, with sufficiently long integration times we obtain good estimate of microcanonical averages of the observables of interest.
Figures \ref{average_curvature_MxR_MxR2} and \ref{fluctuation_curvature_MxR_MxR2} provide the comparison between 
$\langle k_{R}\rangle$ and $\langle\delta^{2}k_{R}\rangle$ for the two manifolds. 
Somewhat unexpectedly these average quantities are found to be practically coincident, thus it is not surprising that the application of the effective scalar model for the JLC equation - recalled above - yields outcomes in close agreement, as shown by Figure \ref{comparison_geo_exp_MxR_MxR2}.

\begin{figure}[H]
	\centering
	\includegraphics[height=10cm, width=10cm,keepaspectratio]{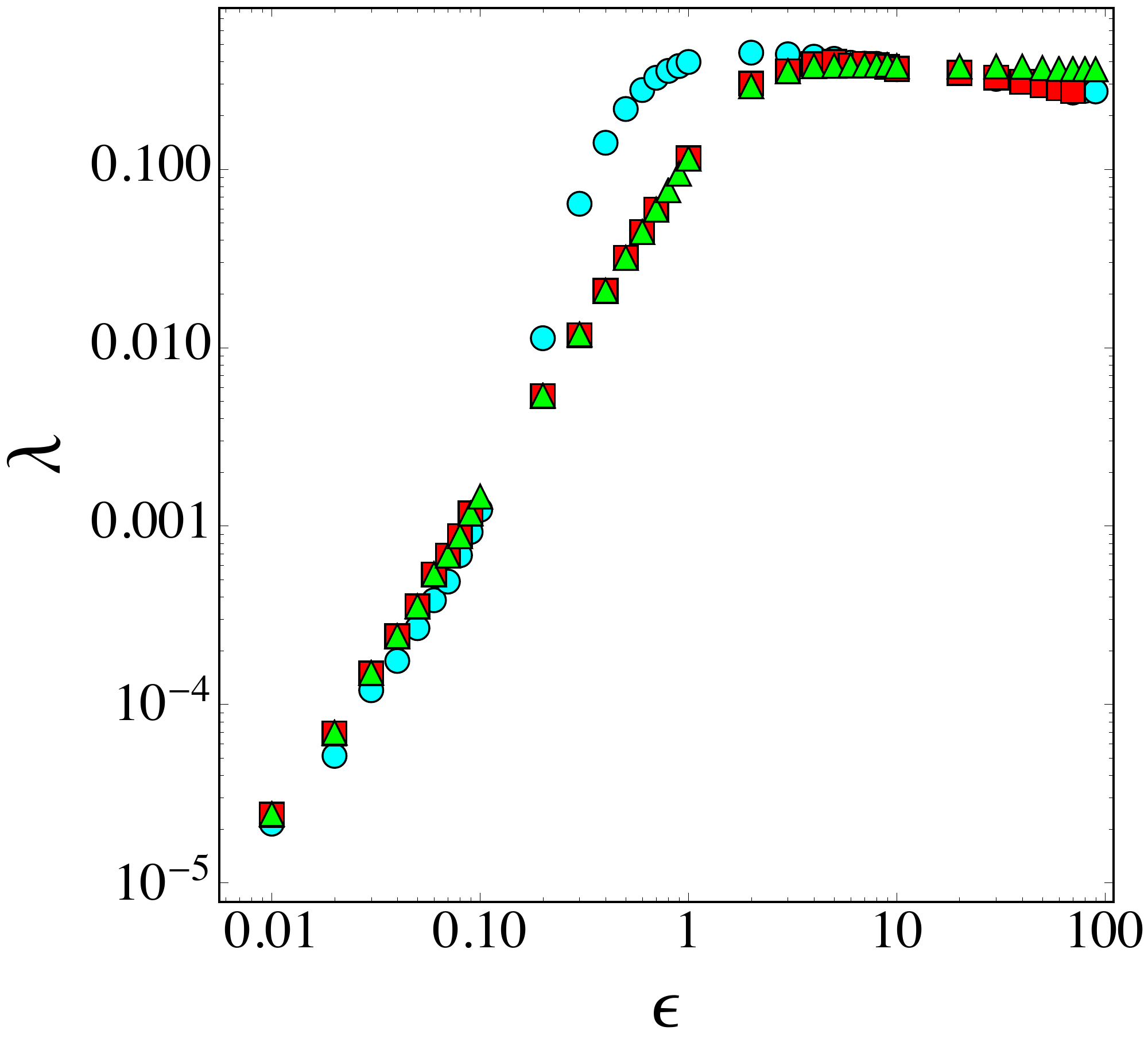}
	\caption{Comparison between the two Geometric Lyapunov Exponents $\lambda_{g_e}$ (red squares), $\lambda_{G_e}$   
	(green triangles) and the standard numerical computation of $\lambda_1$ (cyan circles) vs  energy density $\epsilon$ for a  
	system of $N=150$. Here $\eta = E$.}
	\label{comparison_dintan_geo_exp_MxR_MxR2}
\end{figure}
The comparison among the outcomes  $\lambda_{g_e}(\epsilon)$, $\lambda_{G_e}(\epsilon)$ of the "statistical" formula \eqref{lambdaeisenhart}, and the standard computation of $\lambda_1(\epsilon)$ are displayed in Figure \ref{comparison_dintan_geo_exp_MxR_MxR2}. The discrepancy, observed approximately for  $\epsilon$ in the interval between $0.2$ and $2$, has been given an explanation in  Ref.\cite{CasClePet} where it has been shown that the numerical distribution of the Ricci curvature of ${M\times{\mathbb R}^{2}}$ actually displays a non-vanishing skewness with an excess of negative values with respect to a Gaussian distribution. This information is lost in the effective scalar model for the JLC equation above recalled. 
An \textit{ad hoc} displacement of $\langle k_{R}\rangle$ to empirically account for the excess of negative values of $K_R$ allowed to exactly retrieve the pattern of  $\lambda_1(\epsilon)$ by means of the scalar effective model. 
A-priori the use of $(M\times{\mathbb R},G_{e})$ could have fixed the problem more naturally but, disappointedly, this has not been the case thus calling for an improvement of the effective scalar model, possibly taking into account higher order moments of the Ricci curvature distribution. Finally, it is worth to mention that the potential function of the Hamiltonian \eqref{H_XY_Model} has a large number of critical points $q_c$, that is such that $\nabla V(q)\vert_{q=q_c} = 0$ \cite{book}; near each critical point, in Morse chart one has $V(q) =  V(q_c) -\sum_{i=1}^k q_i^2 + \sum_{i=k+1}^N q_i^2$ where $k$ is the Morse index of a given critical point.
Now, the neighborhoods of critical points are enhancers of chaos because using the expression for $V(q)$ in Morse chart together with $\nabla V(q_c) = 0$, both equations \eqref{dyntanT}  and \eqref{jlcge} diagonalize with $k$ unstable components in proximity of a critical point of index $k$. Morse theory relates critical points of a suitable real valued function (here the potential function) with topological properties of its levels sets, here of equipotential manifolds in configuration space. In other words, the 1D XY model highlights the necessity of taking into account also some topological property of the mechanical manifolds in order to improve the effective scalar model for the JLC equation.

\section{Discussion}
Summarizing, the geometrization of Hamiltonian dynamics within the framework of the configuration space-time equipped with an Eisenhart metric, $(M\times{\mathbb R},G_{e})$, provides a correct distinction of regular and chaotic motions and it is in qualitative and quantitative agreement with the two other geometrization frameworks reported above. As already remarked, the advantage of this framework could be that of an intermediate level of complexity/richness of its geometry with respect to $(M_E,g_J)$ and 
$(M\times{\mathbb R}^{2},g_{e})$ which could be useful in performing more elaborated investigations about the relation between geometry and chaos.

\medskip

Let us conclude with an outlook at a prospective extension to generic dynamical systems of the geometric description of chaos in systems of differential equations
\begin{equation}
\dot x^i = f^i(x^1,\dots,x^N) = f^i(\boldsymbol{x})
\label{dynsys}
\end{equation}
that is, also in the case of dissipative systems.  By differentiation with respect to time of Eq.\eqref{dynsys} we get a new system of equations
\begin{equation}
\ddot x^i = \sum_{j=1}^N \frac{\partial f^i(\boldsymbol{x})}{\partial x^j} \dot x^j =  \sum_{j=1}^N \frac{\partial f^i(\boldsymbol{x})}{\partial x^j}f^j(\boldsymbol{x})
\label{dynsysAdj}
\end{equation}
that can be derived from the Lagrangian function
\begin{equation}
L(\boldsymbol{x},\boldsymbol{\dot x}) = \sum_{i=1}^N [{\dot x}^i - f^i(\boldsymbol{x})]^2
\end{equation}
and the usual Lagrange equations. To this Lagrangian $L(\boldsymbol{x},\boldsymbol{\dot x})$ one associates a metric function 
homogeneous of degree one in the velocities
\begin{equation}
\Lambda(x^a,{\dot x}^a) = L(x^i,{\dot x}^i/{\dot x}^0){\dot x}^0\ , \hskip 0.5truecm  a=0,1,\dots,N;\  i=1,\dots,N
\end{equation}
involving an extra velocity $\dot x^0$; through this metric function a metric tensor expressed as 
\begin{equation}
g_{ab}(\boldsymbol{x},\boldsymbol{\dot x}) = \frac{1}{2}\frac{\partial^2 \Lambda^2}{\partial\dot x^a\partial\dot x^b} 
\end{equation}
provides the tangent bundle of the configuration space of the system \eqref{dynsys} with a Finslerian structure.
The geodesics of this space, minimizing the functional $\int_{\tau_0}^{\tau_1}\Lambda(x^a,{\dot x}^a)d\tau$, are given by \cite{marco,rund}
\begin{equation}
\frac{d^2x^a}{ds^2} + \gamma^a_{bc} (\boldsymbol{x},\boldsymbol{\dot x})\frac{d x^b}{ds}\frac{d x^c}{ds} = 0
\end{equation}
where $\gamma^a_{bc} (\boldsymbol{x},\boldsymbol{\dot x})$ are the connection coefficients derived from the velocity dependent metric $g_{ab} (\boldsymbol{x},\boldsymbol{\dot x})$, and coincide with the solutions of Eqs.\eqref{dynsysAdj}.
Then a geodesic deviation equation is defined also on Finsler manifolds and relates  stability/instability of the geodesics with the curvature properties of the space \cite{marco}.
This approach certainly deserves to be investigated to tackle chaotic dynamics of dissipative systems with the same methodological approach successfully applied to Hamiltonian systems.
\bigskip

\textbf{Acknowledgments}

M.P. participated in this work within the framework of the project MOLINT which has received funding from the Excellence Initiative 
of Aix-Marseille University - A*Midex,  a French “Investissements d’Avenir” programme. 

~
\nocite{*}
\bibliographystyle{plain}
\bibliography{biblio}

\begin{thebibliography}{10}

\bibitem{anosov}
D.~V. Anosov.
\newblock {Geodesic flows on closed Riemannian manifolds with negative
  curvature}.
\newblock {\em Proc. Steklov Math. Inst.}, 90:1--235, 1967.

\bibitem{lapo}
L.~Casetti.
\newblock {Efficient symplectic algorithms for numerical simulations of
  Hamiltonian flows}.
\newblock {\em Physica scripta}, 51(1):29, 1995.

\bibitem{CasClePet}
L.~Casetti, C.~Clementi, and M.~Pettini.
\newblock {Riemannian theory of Hamiltonian chaos and Lyapunov exponents}.
\newblock {\em Physical Review E}, 54(6):5969, 1996.

\bibitem{cerruti1997lyapunov}
M.~Cerruti-Sola, R.~Franzosi, and M.~Pettini.
\newblock {Lyapunov exponents from geodesic spread in configuration space}.
\newblock {\em Physical Review E}, 56(4):4872, 1997.

\bibitem{cerruti1996geometric}
M.~Cerruti-Sola and M.~Pettini.
\newblock {Geometric description of chaos in two-degrees-of-freedom Hamiltonian
  systems}.
\newblock {\em Physical Review E}, 53(1):179, 1996.

\bibitem{cecmar}
C.~Clementi and M.~Pettini.
\newblock {A geometric interpretation of integrable motions}.
\newblock {\em Celestial Mechanics and Dynamical Astronomy}, 84(3):263--281,
  2002.

\bibitem{cuervo2015non}
E.~Cuervo-Reyes and R.~Movassagh.
\newblock {Non-affine geometrization can lead to non-physical instabilities}.
\newblock {\em Journal of Physics A: Mathematical and Theoretical},
  48(7):075101, 2015.

\bibitem{loris}
L.~Di~Cairano, M.~Gori, and M.~Pettini.
\newblock {Coherent Riemannian-geometric description of Hamiltonian order and
  chaos with Jacobi metric}.
\newblock {\em Chaos: An Interdisciplinary Journal of Nonlinear Science},
  29(12):123134, 2019.

\bibitem{Eisenhart}
L.~P. Eisenhart.
\newblock {Dynamical trajectories and geodesics}.
\newblock {\em Annals of Mathematics}, pages 591--606, 1929.

\bibitem{chaos}
J.~Guckenheimer and P.~Holmes.
\newblock {Nonlinear oscillations, dynamical systems and bifurcations of vector
  fields}.
\newblock {\em Appl. Math. Sci. Series}, 42, 1983.

\bibitem{henon}
M.~H{\'e}non and C.~Heiles.
\newblock {The applicability of the third integral of motion: some numerical
  experiments}.
\newblock {\em The Astronomical Journal}, 69:73, 1964.

\bibitem{krylov}
N.~S. Krylov.
\newblock {\em {Works on the foundations of statistical physics}}.
\newblock Princeton Univ. Press, 1979.

\bibitem{lichnerowicz}
A.~Lichnerowicz and T~Teichmann.
\newblock {Th{\'e}ories relativistes de la gravitation et de
  l'{\'e}lectromagn{\'e}tisme}.
\newblock {\em PhT}, 8(10):24, 1955.

\bibitem{chaos1}
A.~J. Lichtenberg and M.~A. Lieberman.
\newblock {\em {Regular and chaotic dynamics}}.
\newblock Springer-Verlag, Berlin, 1992.

\bibitem{JSP}
R.~Livi, M.~Pettini, S.~Ruffo, and A.~Vulpiani.
\newblock {Chaotic behavior in nonlinear Hamiltonian systems and equilibrium
  statistical mechanics}.
\newblock {\em Journal of statistical physics}, 48(3-4):539--559, 1987.

\bibitem{Note1}
The natural and elegant geometric setting of Hamiltonian dynamics is provided
  by symplectic geometry. This geometrical framework is very powerful to study,
  for example, symmetries. However, symplectic manifolds are not endowed with a
  metric, and without a metric we do not know how to measure the distance
  between two nearby phase space trajectories and thus to study their
  stability/instability through the time evolution of such a distance.

\bibitem{marco}
M.~Pettini.
\newblock {Geometrical hints for a nonperturbative approach to Hamiltonian
  dynamics}.
\newblock {\em Physical Review E}, 47(2):828, 1993.

\bibitem{book}
M.~Pettini.
\newblock {\em {Geometry and topology in Hamiltonian dynamics and statistical
  mechanics}}, volume~33.
\newblock Springer Science \& Business Media, 2007.

\bibitem{rick}
M.~Pettini and R.~Valdettaro.
\newblock {On the Riemannian description of chaotic instability in Hamiltonian
  dynamics}.
\newblock {\em Chaos: An Interdisciplinary Journal of Nonlinear Science},
  5(4):646--652, 1995.

\bibitem{poincare}
H.~Poincar{\'e}.
\newblock {\em {Les m{\'e}thodes nouvelles de la m{\'e}canique c{\'e}leste}},
  volume~3.
\newblock Blanchard, Paris, 1987.

\bibitem{rund}
H.~Rund.
\newblock {\em {The differential geometry of Finsler spaces}}, volume 101.
\newblock Springer Science \& Business Media, 2012.

\bibitem{wiggins}
S~Wiggins.
\newblock {Global bifurcations and Chaos}.
\newblock {\em Applied Mathematial Sciences}, 73, 1988.

\end{thebibliography}

\end{document}